\documentclass[article]{emulateapj}
\usepackage{graphicx}
\usepackage{graphics}
\usepackage{natbib}
\usepackage{subfigure}
\usepackage{multirow}
\usepackage{verbatim}
\usepackage[colorlinks=true,linkcolor=blue,urlcolor=blue,citecolor=blue,pdfauthor=derajan,backref=true, bookmarks=false]{hyperref}
\usepackage{amsmath}
\usepackage{setspace}

\usepackage{color}
\usepackage{rotating}
\usepackage{microtype}
\usepackage{etoolbox}
\makeatletter
\patchcmd{\NAT@citex}
  {\@citea\NAT@hyper@{%
     \NAT@nmfmt{\NAT@nm}%
     \hyper@natlinkbreak{\NAT@aysep\NAT@spacechar}{\@citeb\@extra@b@citeb}%
     \NAT@date}}
  {\@citea\NAT@nmfmt{\NAT@nm}%
   \NAT@aysep\NAT@spacechar\NAT@hyper@{\NAT@date}}{}{}
\patchcmd{\NAT@citex}
  {\@citea\NAT@hyper@{%
     \NAT@nmfmt{\NAT@nm}%
     \hyper@natlinkbreak{\NAT@spacechar\NAT@@open\if*#1*\else#1\NAT@spacechar\fi}%
       {\@citeb\@extra@b@citeb}%
     \NAT@date}}
  {\@citea\NAT@nmfmt{\NAT@nm}%
   \NAT@spacechar\NAT@@open\if*#1*\else#1\NAT@spacechar\fi\NAT@hyper@{\NAT@date}}
  {}{}
\makeatother

\providecommand{\adsurl}[1]{\href{#1}{ADS}}

\shortauthors{Lamee, M. et al.}
\begin{document}
\title{Magnetic field disorder and Faraday effects on the polarization of extragalactic radio sources}
\author{Mehdi Lamee  \altaffilmark1 , Lawrence Rudnick  \altaffilmark1, Jamie S. Farnes  \altaffilmark {2,3}, Ettore Carretti  \altaffilmark {4,5}, B. M. Gaensler  \altaffilmark{2,7,8}, Marijke Haverkorn \altaffilmark{3,6} \& Sergio Poppi  \altaffilmark4 }
\affil{\altaffilmark1 Minnesota Institute for Astrophysics, School of Physics and Astronomy, University of Minnesota,
116 Church Street SE, Minneapolis, MN 55455, USA; \href{mailto:lamee@astro.umn.edu}{lamee@astro.umn.edu}}
\affil{\altaffilmark2 Sydney Institute for Astronomy, School of Physics, The University of Sydney, NSW 2006, Australia}\affil{\altaffilmark3 Department of Astrophysics/IMAPP, Radboud University, PO Box 9010, NL-6500 GL Nijmegen, the Netherlands}

\affil{\altaffilmark4 INAF - Osservatorio Astronomico di Cagliari, Via della Scienza 5, 09047 Selargius (CA)}\affil{\altaffilmark5 CSIRO Astronomy and Space Science, PO Box 76, Epping, NSW 1710, Australia}
\affil{\altaffilmark6 Leiden Observatory, Leiden University, PO Box 9513, 2300 RA Leiden, the Netherlands}
\affil{\altaffilmark7 ARC Centre of Excellence for All-sky Astrophysics (CAASTRO), Australia}
\affil{\altaffilmark8 Dunlap Institute for Astronomy and Astrophysics, University of Toronto, ON, M5S 3H4, Canada}

\begin{abstract}
We present a polarization catalog of 533 extragalactic radio sources with 2.3 GHz total intensity above 420 mJy from the S-band Polarization All Sky Survey, S-PASS, with corresponding 1.4 GHz polarization information from the NRAO VLA Sky Survey, NVSS. We studied selection effects and found that fractional polarization, $\pi$, of radio objects at both wavelengths depends on the spectral index, source magnetic field disorder, source size and depolarization.  The relationship between depolarization, spectrum and size shows that depolarization occurs primarily in the source vicinity. The median $\pi_{2.3}$ of resolved objects in NVSS is approximately two times larger than that of unresolved sources.  Sources with little depolarization are $\sim2$ times more polarized than both highly depolarized and re-polarized sources. This indicates that intrinsic magnetic field disorder is the dominant mechanism responsible for the observed low fractional polarization of radio sources at high frequencies. We predict that number counts from polarization surveys will be similar at 1.4 GHz and at 2.3 GHz, for fixed sensitivity, although $\sim$10\% of all sources may be currently missing because of strong depolarization. Objects with $\pi_{1.4}\approx \pi_{2.3} \ge 4\%$ typically have simple Faraday structures, so are most useful for background samples. Almost half of flat spectrum ($\alpha \ge -0.5$) and $\sim$25\% of steep spectrum objects are re-polarized. Steep spectrum, depolarized sources show a weak negative correlation of depolarization with redshift in the range 0 $<$ z $<$ 2.3. Previous non-detections of redshift evolution are likely due the inclusion of re-polarized sources as well.
   
\end{abstract}

\section{Introduction}
There are many open questions regarding the strength and geometry of the magnetic field in radio galaxies and their relation to other properties of the radio source. The observed degree of polarization depends on the intrinsic properties, such as the regularity and orientation of the source magnetic fields as well as the Faraday effects from the intervening regions of ionized gas along the line of sight. The largest current sample of polarized sources is the NRAO/VLA all sky survey, NVSS, at 1.4 GHz \citep{1998AJ....115.1693C}. It shows that the majority of extragalactic radio sources are only a few percent polarized. Polarization studies of small samples of extragalactic radio sources at other frequencies also show a similar weak average polarization, and suggest the fractional polarization increases at frequencies higher than 1.4 GHz \citep[e.g.][]{2009A&A...502...61M}.  It is not clear which mechanism is dominant in reducing the fractional polarization at lower frequencies and depolarizing the sources, although several models have been suggested \citep{1966MNRAS.133...67B,1991MNRAS.250..726T,1998MNRAS.299..189S,2008A&A...487..865R,2015MNRAS.450.3579S}. 

One key cause for depolarization is Faraday rotation, which can be characterized to first order by a change in the angle of the linear polarization:
\begin{equation}  
\Delta \chi=\left(0.812 \int \frac{n_e{\bf B}}{(1+z)^2}\cdot \frac{d{\bf l}}{dz} \,dz\right) \lambda^2  \equiv \phi \lambda^2 
\end{equation}
where $\Delta \chi$ is the amount of the rotation of the polarization vector in rad, $\lambda$ is the observation wavelength in m, $z$ is the redshift of the Faraday screen, ${\bf B}$ is the ionized medium magnetic field vector in $\mu$G, $n_e$ is the number density of electrons in the medium in cm$^{-3}$ and $\,d{\bf l}$ is the distance element along the line of sight in pc. The term in parentheses is called the Faraday depth, $\phi$. For a single line of sight through a thin ionized screen, this is equivalent to the rotation measure, $\textrm{RM}$, defined by $\textrm{RM} \equiv \frac{\Delta \chi}{\Delta \lambda^2}$ which can be measured observationally. 

Different lines of sight to the source all within the observing beam can have different values of $\phi$. Typically, this progressively depolarizes the source at longer wavelengths, but it can also lead to constructive interference and re-polarization, i.e., higher fractional polarizations at longer wavelengths.
There are at least three separate possible Faraday screens with different $\textrm{RM}$ distributions along the line of sight: the Galactic component, intervening extragalactic ionized gas, and material local to the source. Multiple studies such as  \cite{2005MNRAS.359.1456G,2008ApJ...676...70K,2010MNRAS.409L..99S,2012ApJ...761..144B,2012arXiv1209.1438H,2013ApJ...771..105B,2014ApJ...795...63F,2014MNRAS.444..700B,2014PASJ...66...65A,2015aska.confE.114V,2015arXiv150900747V} have tried to identify and distinguish these separate components and study the evolution of the magnetic field of galaxies through cosmic time. When many lines of sight each have independent single Faraday depths,  this problem is approached statistically. 

Another long standing puzzle is the anti-correlation between the total intensity of radio sources and their degree of polarization, as observed by many groups such as \cite{2002A&A...396..463M}, \cite{2004MNRAS.349.1267T}, \cite{2006MNRAS.371..898S}, \cite{2007ApJ...666..201T}, \cite{2010ApJ...714.1689G}, \cite{2010MNRAS.402.2792S} and \cite{2014ApJ...787...99S}. The physical nature of this relation has been a mystery for almost a decade, and is confused by the dependency on other source properties. \cite{2010ApJ...714.1689G} found that most of their highly polarized sources are steep spectrum, show signs of resolved structure on arc-second scales, and are lobe dominated. However, they found no further correlation between the spectral index and fractional polarization. The anti-correlation between total intensity and fractional polarization seems to become weak for very faint objects with 1.4 GHz total intensities between 0.5 mJy $< I <$ 5 mJy as suggested in \cite{2014ApJ...785...45R}, based on a small sample of polarized radio galaxies in the GOODS-N field \citep{2010ApJS..188..178M}.  Recently, \cite{2015arXiv150406679O} studied a sample of 796 radio-loud AGNs with $z < 0.7$. They found that low-excitation radio galaxies have a wide range of fractional polarizations up to $\sim$ 30 \%, and are more numerous at faint Stokes I flux densities while high-excitation radio galaxies are limited to polarization degrees less than 15\%. They suggest that the ambient gas density and magnetic fields local to the radio source might be responsible for the difference. 
Using WISE colors, \cite{2014MNRAS.444..700B}  suggested that the observed anti-correlation primarily reflects the difference between infrared AGN and star-dominated populations.  

Large samples of polarization data at multiple frequencies are required to understand the magnetic field structures and depolarization mechanisms responsible for the low observed polarization fractions.  \cite{2013ApJ...771..105B} have showed the polarization fraction of compact sources decreases significantly at 189 MHz compared to 1.4 GHz. They studied a sample of 137 sources brighter than 4 mJy and only detected one polarized source with probably a depolarization mechanism intrinsic to the source. Recently, \cite{2014ApJS..212...15F} used the \cite{2009ApJ...702.1230T} (hereafter TSS09) catalog, and assembled polarization spectral energy distributions for 951 highly polarized extragalactic sources over the broad frequency range, 0.4 GHz to 100 GHz. They showed that objects with flat spectra in total intensity have complicated polarization spectral energy distributions (SEDs), and are mostly re-polarized somewhere in the spectrum, while steep spectrum sources show higher average depolarization. As a result, they claimed that the dominant source of depolarization should be the local environment of the source, since the spectral index is an intrinsic property of these highly polarized sources. The current work follows up on their discovery, using a sample selected only on the basis of total intensity at 2.3 GHz.

In this work, we use the data from the S-PASS survey,  conducted by the Australian Parkes single dish radio telescope at 2.3 GHz. We cross match the data with the NVSS catalog and generate a new independent depolarization catalog of bright extragalactic radio sources. Unlike other polarization studies such as \cite{2014ApJS..212...15F} and \citet{2012arXiv1209.1438H} our catalog is not selected based on high polarized intensity which enables us to include objects with low fractional polarizations as well. We study the evolution and possible correlation between quantities such as depolarization, spectral indices and $\textrm{RM}$s. We will tackle the nature of the well-known observed anti-correlation between total intensity and fractional polarization as well as the origin of the dominant component of depolarization. Section \ref{sec:obs} presents the 1.4 GHz and 2.3 GHz observations. Section \ref{sec:mapanalysis} explains the steps in our analysis of the S-PASS total intensity and polarization maps as well as the cross matching with the NVSS catalog. In Section \ref{quantities} we derive quantities such as spectral index, residual rotation measure, fractional polarization and depolarization. The main results and their implications are discussed in sections \ref{result} and \ref{discussion} respectively. At the end, Section \ref{summary} summarizes the main findings and conclusions.     

Throughout this paper we employ the $\Lambda$CDM cosmology with parameters of H$_0=70$ km.s$^{-1}$Mpc$^{-1}$, $\Omega_m=0.3$ and $\Omega_{\Lambda}=0.7$. 

\section{Observations} \label{sec:obs}
\subsection{The 2.3 GHz Data}\label{spass}
The S-PASS is a project to map the southern sky at Dec $<-1.0$ deg in total intensity and linear polarization. The observations were conducted with the 64-m Parkes Radio Telescope, NSW Australia. A description of S-PASS is given in \cite{2013Natur.493...66C} and \cite{2010ASPC..438..276C}; here we report a summary of the main details.

The S-band receiver used is a circular polarization package with  system temperature T$_{sys}$ = 20 K, and beam width FHWM= 8.9 arcmin at 2300 MHz. Data were collected with the digital correlator Digital Filter Banks mark 3 (DFB3) recording the two autocorrelation products (RR* and LL*) and their complex cross-correlation (RL*). The sources PKS B1934-638 and PKS B0407-658 were used for flux density calibration and PKS B0043-424 for polarization calibration. After Radio Frequency Interference (RFI) flagging, frequency channels were binned together covering the ranges 2176-2216 and 2256-2400 MHz, for an effective central frequency of 2307 MHz and bandwidth of 184 MHz. 

As described in \cite{2010ASPC..438..276C}, the observing strategy is based on long azimuth scans taken at the elevation of the south celestial pole at Parkes covering the entire Dec range (-89 deg to -1 deg) in each scan. For the current work, the spatial large scale component has been removed from each Stokes parameter, applying a high pass spatial filter to optimize for compact source finding and analysis. A median filter with a window of 45 arc-min was used. The final product was a set of 15$\times$15 deg$^2$ zenithal projection maps covering the entire sky observed by S-PASS. Final maps are convolved to a beam of FWHM = 10.75 arcmin. Stokes I , Q , and U sensitivity is better than 1.0 mJy beam$^{-1}$. Details of scanning strategy, map-making, and final maps are in \cite{2010ASPC..438..276C} and \cite{2013Natur.493...66C}  and will be presented in full details in a forthcoming 
paper (Carretti et al. 2016, in preparation). The confusion limit is 6 mJy in Stokes I \citep{2013MNRAS.430.1414C}  and much lower in polarization (average polarization fraction in compact sources is around 2\%, see this work). The instrumental polarization leakage is 0.4\% on-axis \citep{2010ASPC..438..276C} and less than 1.5\% off-axis.

\subsection{The 1.4 GHz Data}

The NVSS is a 1.4 GHz radio survey with the Very Large Array (VLA) covering the entire sky north of -40 degrees declination at a resolution of 45 arcsec (FWHM). The rms brightness fluctuations are approximately uniform across the sky at $\sim$0.45 mJy per beam in Stokes I and $\sim$0.29 mJy per beam in Stokes Q and U. The astrometry is accurate to within $<1$ arcsec for point sources with flux densities $>15$ mJy, and to $<7$ arcsec for the faintest detectable sources ($\sim$2.3 mJy in Stokes I). The survey has a completeness limit of 2.5 mJy, which resulted in a catalog of over 1.8 million discrete sources in Stokes I. More details about the NVSS can be found in \cite{1998AJ....115.1693C}.

\section{Creating the new sample}
\subsection{Cross-matching and selection criteria}\label{sec:mapanalysis}
We first attempted to construct a joint S-PASS/NVSS catalog using
NVSS I,Q, and U images convolved to the processed S-PASS resolution of
$\sim$11'.  However, upon convolution, the resulting NVSS images were
very heavily mottled because of the lack of short interferometer
baselines, and the noise level increased dramatically above the full
resolution images. We therefore followed an alternative approach, viz.,
measuring the contributions of all individual NVSS sources at the
position of each NVSS source, as described further below.  There are
rare situations where very-low level diffuse NVSS emission could also have
contributed significantly to the S-PASS flux \citetext{e.g, cluster halos, \citealt{2001ApJ...548..639K}}, and would be missed by
our procedure, but this very minor possible contribution to our strong total intensity sources has been ignored.

We constructed the initial  S-PASS catalog by searching the S-PASS maps at the position
of all NVSS sources with $I_\mathrm{NV}~>~$10~mJy in the overlap region between
the two surveys, and fitting Gaussian functions to the S-PASS total intensity
images.  For sources with a spectral index of -0.7 (-0.3)  this would
correspond to a 4(5) $\sigma$ detection in S-PASS.  However, in order to
have adequate sensitivity  to sources with low fractional polarizations in S-PASS, we
adopted a much higher threshold of $I_\mathrm{SP} >$ 420~mJy for the catalog.
Duplicate sources were eliminated.

  Additional sources were eliminated from the catalog if they had
either of these data quality issues:\\
a) Excess noise ($>$0.75~mJy per beam rms, 1.5 $\times$ the mode calculated in bins of 0.01 mJy) in the 7.5' -
11.25' annulus around the total intensity NVSS source;\\
b) Excess noise ($>$3~mJy per beam rms. 2$\times$ the average rms value) in the
45'-90' annulus in either Q or U maps in S-PASS.\\ We verified by visual inspection that the above selection criteria have successfully eliminated the NVSS
and S-PASS regions with instrumental artifacts.

At the processed S-PASS resolution of $\sim$11', many sources identified by the
above procedure are actually blends of multiple NVSS sources.  In order
to derive meaningful information from the sample, we therefore
needed to eliminate sources with significant contributions from
blending.   To do this, we defined a search radius of 16' (i.e., to the 3.5$\sigma$, 2$\times 10^{-3}$ level of the
S-PASS beam) around each S-PASS source, and calculated  the I,Q, and U contributions of each NVSS source
(with $I_\mathrm{NV}>$10 mJy) at the position
of the S-PASS source.  Thus, for the NVSS portion of the catalog, we have
two values for each Stokes parameter: X$_{Ntarget}$, the flux (I,Q,
or U)  of the NVSS source with the largest Stokes I contribution at the
S-PASS position, and X$_{Ncont}$, the I,Q, or U flux from all other
NVSS sources within the 16' search radius, scaled  by their
distance from the S-PASS peak position using a Gaussian kernel representing the S-PASS beam.  The final values for comparison
with S-PASS are then X$_{Ntotal} \equiv$ X$_{Ntarget}$ + X$_{Ncont}$.

\begin{figure}[h]
\centering
\includegraphics[scale=0.51]{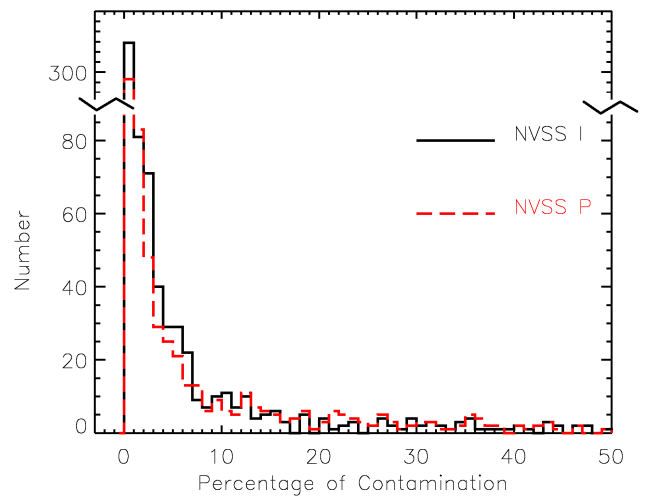}
\caption{ Distributions of the percentage of contamination in the NVSS total intensity, $100\times\frac{I_{cont}}{I_{target}}$, in black solid line and polarization flux density, $100\times\frac{P_{cont}}{P_{target}}$ in red dashed line are shown. The catalog only contains sources with  $\frac{P_{cont}}{P_{target}} < 0.1$.
}
\label{contI2}
\end{figure}

Figure \ref{contI2} shows the distribution of the percent contamination of the target source in NVSS total intensity, $\frac{I_{\text{cont}}}{I_{\text{target}}}$, and polarization, $\frac{P_{\text{cont}}}{P_{\text{target}}}$. We then adopted a 10\% polarization contamination threshold, and only selected sources with $\frac{P_{\text{cont}}}{P_{\text{target}}} < 0.1$.     

The joint S-PASS/NVSS catalog contains 533 sources meeting all of the
above criteria.  A description of the biases that could result from our
contamination threshold is discussed in Section \ref{bias}.

\subsection{Derived quantities}\label{quantities}
\subsubsection{NVSS and S-PASS polarized flux density, fractional polarization and depolarization}
We calculated the polarization intensity (averaged over the entire bandwidth) for the NVSS and S-PASS surveys separately. The effect of bandwidth depolarization is discussed in section \ref{bwdepol}. We used Stokes $Q$ and $U$ to calculate the polarized intensity, $P$, in both NVSS and S-PASS as following:
\begin{equation}
P=\sqrt{Q^2+U^2}
\end{equation}
where for NVSS the $Q$ and $U$ include both the target and contamination flux density,  $Q=Q_{\text{target}}+Q_{\text{cont}}$ and $U=U_{\text{target}}+U_{\text{cont}}$.
The bias corrected polarized flux density, $P_{bc}$, is approximated as follows:
\begin{equation}
P_{\text{bc}}=\sqrt{Q^2+U^2-\sigma_{p}^2-\sigma_{\text{cont}}^2}
\end{equation}
where $\sigma_{p}$ is the global rms of $U$ or $Q$ maps ($\sigma^{NV}_{U,Q} \approx 0.3$ mJy per beam and $\sigma^{SP}_{U,Q} \approx 1.7$ mJy per 3-arcmin pixel), measured through the entire $Q$ and $U$ maps, and $\sigma_{\text{cont}}$ is the total contribution of the contaminant apertures rms noise to the bias in NVSS, scaled for their separation from the target.

We also calculated the fractional polarization, $\pi$, 
\begin{equation}
\pi_\mathrm{SP}=\frac{P^{SP}_{\text{bc}}}{I_\mathrm{SP}}
\end{equation}
\begin{equation}
\pi_\mathrm{NV}=\frac{P^{NV}_{\text{bc}}}{I_\mathrm{NV}}
\end{equation}
where the NVSS total intensity is equal to the target plus the contamination flux density, $I_\mathrm{NV}=I_{\text{target}}+I_{\text{cont}}$.
The NVSS residual instrumental polarization percentage peaks at $\epsilon_\mathrm{NV} \approx 0.12\%$ for a sample of strong and unpolarized sources \citep{1998AJ....115.1693C}. We used this value as a cutoff for the NVSS fractional polarization; for any sources below this threshold we report upper limits as the maximum of  $\left(\frac{3\sigma_p}{I}, \epsilon_\mathrm{NV} \right)$. 

\begin{figure}[h]
\centering
\includegraphics[scale=0.51]{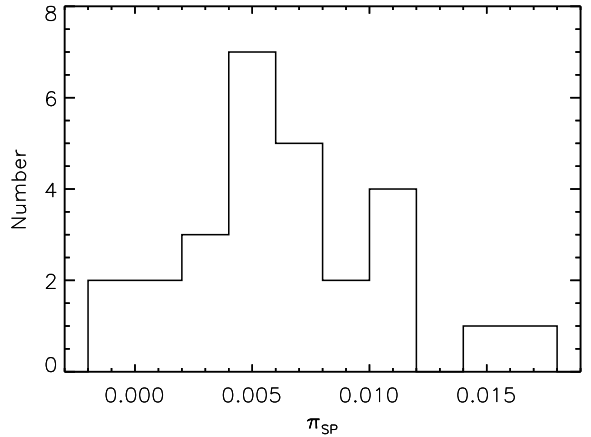}
\caption{ Distribution of S-PASS fractional polarization for 27 objects with $\pi_\mathrm{NV} < \epsilon_\mathrm{NV} $.}
\label{fig:sprespi}
\end{figure}

To estimate the S-PASS residual instrumental polarization we selected the 27 objects with $\pi_\mathrm{NV} < 0.12\%$ in our final sample, and plotted the distribution of their $\pi_\mathrm{SP}$ values (Figure \ref{fig:sprespi}). The median of the distribution, $\bar{\pi}_\mathrm{SP}=0.55\%$, which we assumed to be a good estimator of the S-PASS residual instrumental polarization percentage, $\epsilon_\mathrm{SP}$. Note that if the residual instrumental polarizations were zero, then the rms noise of 1.7 mJy per beam would result in much smaller fractional polarizations than $0.55\%$ for the 27 mentioned objects. On the other hand, objects with $\pi_\mathrm{NV} < 0.12\%$ can potentially be more polarized at higher frequencies, so we could be overestimating the instrumental contribution. We ignored this possibility, and chose the more conservative approach of assuming $\pi_\mathrm{SP}=0.55\%$ is only due to instrument leakage.

Out of $533$ objects, 416 objects are successfully detected ($P_{\text{bc}} > 3\sigma_p$ and $\pi > \epsilon$) in both NVSS and S-PASS polarized flux densities. There are 90 sources that are not detected in polarization in S-PASS but are detected in NVSS, whereas 12 objects with no detection in NVSS polarization are detected in S-PASS. There are 15 objects that do not have polarization above our threshold in either survey.

The depolarization, D,  is defined to be the ratio between S-PASS and NVSS fractional polarizations:
\begin{equation}
D \equiv \frac{\pi_\mathrm{SP}}{\pi_\mathrm{NV}}
\end{equation}
We calculated the depolarization of all objects with $3\sigma_p$ polarization detection and $\pi > \epsilon$ in both S-PASS and NVSS. Upper/lower limits on $D$ are also calculated for sources as appropriate.
\subsubsection{Polarization angle and rotation measure}
Assuming that the contaminating sources have very little impact on the polarization angle of the target source, we used NVSS and S-PASS $Q$ and $U$ flux densities to derive the polarization angles $\chi_\mathrm{NV}$ and $\chi_\mathrm{SP}$; where
\begin{equation}
\chi=\frac{1}{2}\tan^{-1}\frac{U}{Q}
\end{equation} 
These angles are used to estimate the amount of the rotation measure,  $\textrm{RM}_\mathrm{NS}$, between the NVSS and S-PASS. The median uncertainty on the derived rotation measures is on the order of 1.6 rad m$^{-2}$.

The polarization angle can be wrapped by a positive or negative integer coefficient, $n$, of $\pi$ radians from the true angle, the so-called $n\pi$ ambiguity. In this case, the true rotation measure is  $\textrm{RM}_\mathrm{NS}=\textrm{RM}_0 \pm n\pi/\lambda^2$ rad m$^{-2}$. 

We used the TSS09 rotation measure catalog ( $\textrm{RM}_\mathrm{T}$) to fix $n$ by minimizing the absolute values of $\Delta \textrm{RM}$, where $\Delta\textrm{RM} \equiv \textrm{RM}_\mathrm{T}-\textrm{RM}_\mathrm{NS}-n\pi/\lambda^2$ for the 364 sources in common. These are not necessarily the correct $\textrm{RM}$s, since TSS09 has its own $n\pi$ ambiguity of 653 rad m$^{-2}$, while this ambiguity for  $\textrm{RM}_\mathrm{NS}$ is about 108 rad m$^{-2}$. However, they provide the most conservative estimate of $\Delta \textrm{RM}$, the inferred non-linearity in the Faraday rotation as a function of $\lambda$. The parameter $n$ took values of $-1, 0, 1$ for all objects except one with $n=-2$. 

Note that, including the polarization contamination and recalculating the RMs based on the two NVSS sub-bands could introduce offsets as large as 42 rad m$^{-2}$. As a result, using the uncontaminated NVSS $\textrm{RM}_\mathrm{T}$ is appropriate.
\subsubsection{Bandwidth depolarization}\label{bwdepol}
\begin{figure}[h]
\centering
\includegraphics[scale=0.51]{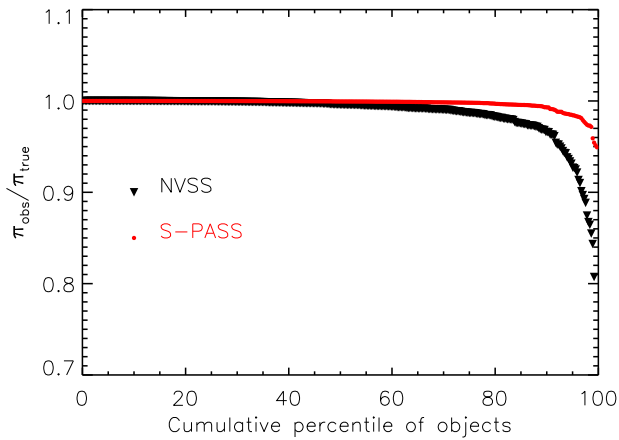}
\caption{The ratio of the observed and true fractional polarizations, $\pi_\mathrm{obs}/\pi_\mathrm{true}$, based on the NVSS and S-PASS bandwidth depolarizations is shown for 364 objects as a function of the cumulative percentile. Only 3\% of objects in our sample experience NVSS bandwidth depolarization which results in $\pi_\mathrm{obs}/\pi_\mathrm{true}$ smaller than 0.9.}
\label{fig:BWdep2}
\end{figure}

When the $\textrm{RM}$ is high, the rotation of the polarization angle across a fixed bandwidth reduces the net degree of polarization, which is called bandwidth depolarization.  To evaluate the importance of this effect for our sample, we used the 364 sources overlapping with TSS09. We predicted  the NVSS and S-PASS bandwidth depolarizations for our objects based on the measured TSS09  $\textrm{RM}_\mathrm{T}$ and our  $\textrm{RM}_\mathrm{NS}$, respectively.  As shown in Figure \ref{fig:BWdep2} the ratio between the observed fractional polarization and the true degree of polarization $\pi_\mathrm{obs}/\pi_\mathrm{true}$ never gets smaller than 0.95 for S-PASS, and only 3\% of objects have NVSS $\pi_\mathrm{obs}/\pi_\mathrm{true}$ smaller than 0.9. The median $\pi_\mathrm{obs}/\pi_\mathrm{true}$ for both S-PASS and NVSS are 0.999 and 0.996 respectively, and therefore, bandwidth depolarization will not affect our analysis throughout this work. 

\subsubsection{Spectral index}
\begin{figure}[h]
\centering
\includegraphics[scale=0.51]{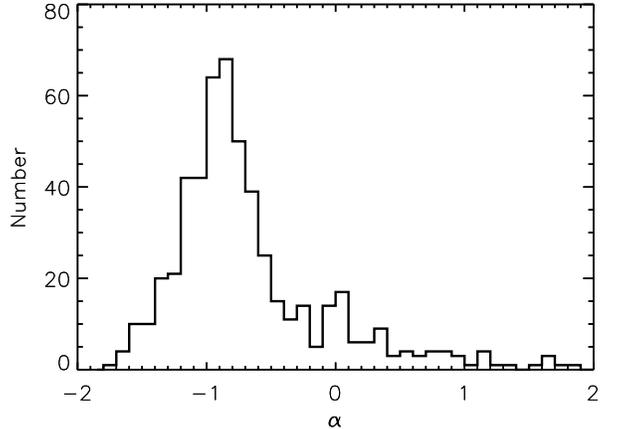}
\caption{ Distribution of spectral indices,$\alpha$, calculated based on NVSS and S-PASS total intensities. The median spectral index is $\bar{\alpha} \approx -0.83$.}
\label{fig:sphst}
\end{figure}

We used I$_{NV}$  and peak S-PASS  (11' beam) intensities, and calculated the power law spectral index, $\alpha$, where
$I \propto \nu^{\alpha}$. Figure \ref{fig:sphst} shows the distribution of spectral indices for our 533 objects. The median is $\bar\alpha\sim -0.83$. The contaminating flux contributing to the NVSS intensities can be a small source of uncertainty in the calculated spectral indices; we estimated its median to be $\sigma_{\alpha,Cont} \sim 0.01$ while total uncertainties on the derived spectral indices has median value of $\sigma_{\alpha,Tot}=0.05$.  
\subsubsection{Surface area of the object}
We used the NVSS catalog de-convolved minor, $\theta_m$, and major $\theta_J$ axes of the target object, and calculated the effective area, $A$ as follows.:
\begin{equation}
A \equiv \frac{1}{4}\pi\theta_m\theta_J
\end{equation} 
One must note that almost all sources remain unresolved in S-PASS due to the very large beam size.
\subsubsection{Uncertainties}
We used the measured local rms values as uncertainties of the NVSS $Q$, $U$ and S-PASS $I$, $Q$ and $U$ flux densities. The uncertainty of the NVSS total intensities are extracted from the NVSS catalog. Error propagation is used to approximate the uncertainty on all the other derived quantities such as polarized flux density and rotation measure. We note that \cite{2011ApJ...726....4S} showed that the rotation measure uncertainties reported in the TSS09 catalog might be underestimated. As a result, we multiplied all the  $\textrm{RM}_\mathrm{T}$ uncertainties by 1.22 as described in  \cite{2011ApJ...726....4S}.   

\subsection{Selection Bias}\label{bias}
We do not select objects based on their polarization intensities or fractional polarizations. However, we apply a threshold cut on the contribution of polarized contaminants. There is a higher probability for objects with low polarized intensity, either intrinsic or due to depolarization, to suffer from contaminating neighbors, and to be dropped from our final sample. To investigate a possible missing population, we compared two different sub-samples  a) sources in our catalog with 
$\frac{P_{\text{cont}}}{P_{\text{target}}} <$ 0.1 (533 sources, 416 detected in both NVSS and S-PASS) and b) objects rejected from our catalog with 0.1 $\le \frac{P_{\text{cont}}}{P_{\text{target}}} <$ 0.25 (75 sources, 40 detected in both NVSS and S-PASS).   

\begin{figure}[h]
\centering
\includegraphics[scale=0.51]{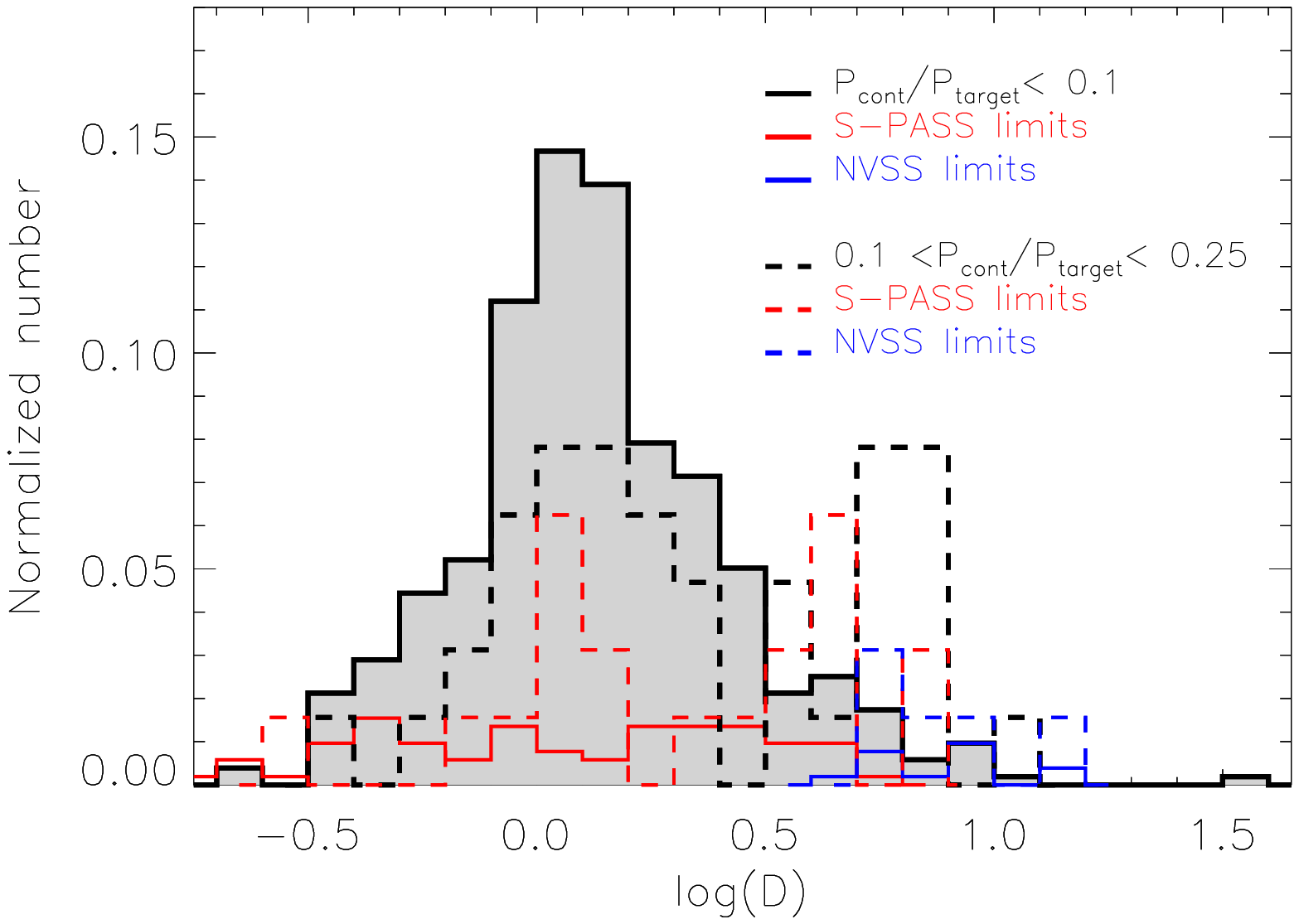}
\includegraphics[scale=0.51]{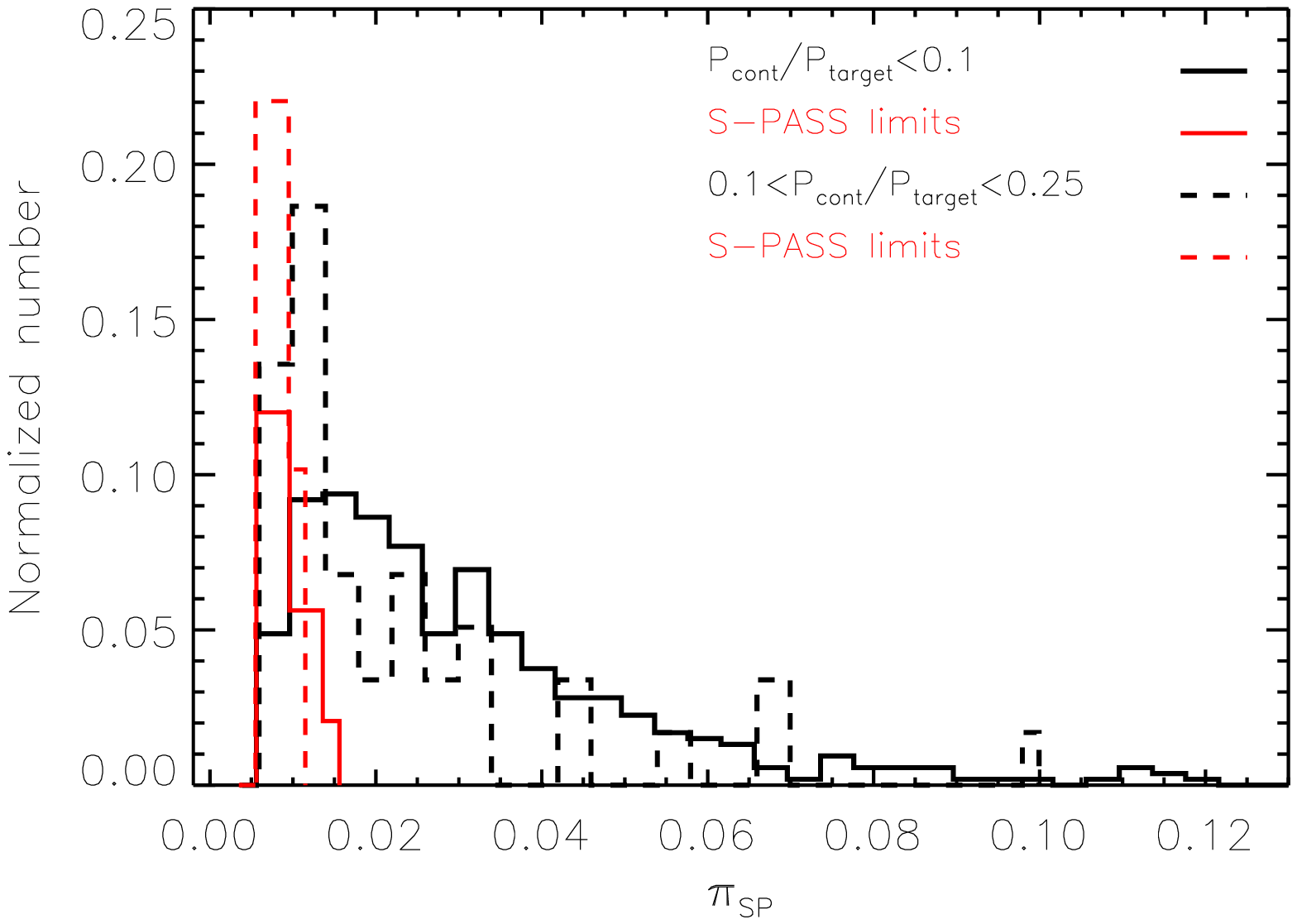}
\caption{Distribution of $\log(D)$ (top) and 2.3 GHz fractional polarization $\pi_\mathrm{SP}$ (bottom) of objects with $\frac{P_{\text{cont}}}{P_{\text{target}}} <$ 0.1 and 0.1 $\le \frac{P_{\text{cont}}}{P_{\text{target}}} <$ 0.25 are shown with solid and dashed lines respectively. Black, red and blue colors represent objects with detection, upper limits in S-PASS and upper limits in NVSS polarizations. The area under the $\log(D)$ histogram of objects with $\frac{P_{\text{cont}}}{P_{\text{target}}} <$ 0.1 and detected polarization flux densities is colored in gray for clarity. }
\label{polcont}
\end{figure}

We compared the fractional polarization and the depolarization properties of these two sub-samples.  If we were \emph{not} creating a selection bias, then they should have similar properties. Figure \ref{polcont} shows the results.
Objects with larger polarization contamination have on average lower 2.3 GHz fractional polarization (median $\bar \pi_\mathrm{SP} = 1.5\%$) while less contaminated sources have $\bar \pi_\mathrm{SP}= 2.5\%$.  Moreover, the fraction of sources with $\pi_\mathrm{SP} < 1\%$ is 2.5 times higher (50\%) among objects with 0.1 $\le \frac{P_{\text{cont}}}{P_{\text{target}}} <$ 0.25 than sources with $\frac{P_{\text{cont}}}{P_{\text{target}}} <$ 0.1.  The Spearman rank test between $D$ and $\frac{P_{\text{cont}}}{P_{\text{target}}}$ with $r= 0.22$ and $p<0.00001$ rejects the null hypothesis of no correlation.  Thus, we are likely to be missing a population of highly depolarized sources.

Figure \ref{polcont} suggests that around 30\% of sources with polarized contamination   0.1 $\le \frac{P_{\text{cont}}}{P_{\text{target}}} <$ 0.25 have depolarizations $\log(D)>0.47$. Assuming this fraction is also valid for sources with contaminations larger than 25\% we estimate that we have missed $\sim 50$ depolarized objects in our final sample due to the polarized contamination threshold cut.  

Therefore, our final sample of 533 sources is missing a population ($\sim 50$ objects) of heavily depolarized sources due to our contamination threshold cut. However, we can not correct for such an effect since the amount of contamination in our 2.3 GHz polarization intensities can not be measured. As a result, one should treat the number of depolarized sources in our sample as a strong lower limit and consider this in interpreting all the other related conclusions.    

In addition, it is possible that our total intensity and polarization contamination thresholds have resulted in a bias toward less dense regions of the sky. We measured the surface number density of the contaminating neighbors in our sample and the parent NVSS--S-PASS overlap sample with $I_\textrm{NV} > 10$ mJy. We used the same aperture with a radius of 16 arcmin and found that the contaminant surface number density in our final sample ($4\times 10^{-3}$ arcmin$^{-2}$) is on average 20\% less than our parent sample ($5\times 10^{-3}$ arcmin$^{-2}$). It is unlikely that the results of this work are affected by such a bias.
\subsection{Statistical tests}\label{stat}
Throughout this work we adopted two nonparametric statistical tests. We calculated the Spearman rank correlation coefficient ($r_s$) to measure the strength of any possible correlation. The two-sample Kolmogorov-Smirnov (KS) test is also used to check the null hypothesis that two sub-samples, divided by a parameter of interest,  are drawn from the same parent distribution.   The significance of each test is estimated by performing bootstrap sampling simulations and constructing $10^5$ random samples from the initial distribution. We have assigned two-tail p-values based on the results of our simulations. 

Table \ref{table2} summarizes the result of all the statistical tests performed in this work. In the case of a single hypothesis test we would reject the null hypothesis if the p-value $\le0.01$. However,  we have performed a total 90 tests, counting both KS and Spearman. To avoid the multiple hypothesis testing problem, we  adopted the Bonferroni correction as discussed in \cite{2016arXiv160203765A} and chose a conservative significance level threshold of p-value $\le 10^{-4}$. We therefore rejected the null hypothesis of the KS or the Spearman rank tests only if the corresponding p-value is less than or equal to $10^{-4}$.    

In addition, to test the robustness of correlations with p-value less than $10^{-4}$ and to identify any possible influence of the total intensity and polarization contaminations on the results we repeated the relevant statistical tests on smaller (by a factor of $\sim$0.4) but clean samples of objects with less than 1\% contamination.  Although the strength of some correlations became stronger or did not change, their p-values increased up to $2\times 10^{-3}$ due to the much smaller sample size. We therefore, adopted the robustness probability p$_\textrm{robust}$ of $2\times 10^{-3}$ as a second threshold and treated the correlations with original p-value$\le10^{-4}$ and $2\times 10^{-3} < $p$_\textrm{robust}< 0.05$ as suggestive relations only, and did not draw any conclusion based on them. These are marked in Table \ref{table2} for completeness and potentially future work. The correlations with original p-value$ <10^{-4}$ but p$_\textrm{robust}> 0.05$ are rejected.  

\section{Results}\label{result}
We have derived a polarization catalog of 533 extragalactic radio sources, which can be downloaded for public use through the VizieR catalog access tool. The description of the entries in the online catalog is listed in table \ref{table3}.

\tabcolsep=0.1cm
\tabletypesize{\scriptsize}
\begin{deluxetable}{ll}[h]
\tablecolumns{2}
\tablewidth{85mm}
\tablecaption{Description of the entries in the online catalog.  \label{table3}}
\tablehead{
\colhead{{Column index}} & \colhead{Description} }

1 & NVSS name tag \\
2 & NVSS RA in decimal degrees (J2000) \\
3 & NVSS Dec in decimal degrees (J2000) \\
4 & Galactic longitude \\
5 & Galactic latitude \\
6 & NVSS total intensity, $I_\mathrm{NV}$ \\
7 & Uncertainty on the $I_\mathrm{NV}$ \\
8 & NVSS polarized intensity \\
9 & Uncertainty on the NVSS polarized intensity \\
10 & NVSS fractional polarization, $\pi_\mathrm{NV}$\\
11 & Uncertainty on the $\pi_\mathrm{NV}$ \\
12 & Upper limit flag on the $\pi_\mathrm{NV}$  \\
13 & NVSS polarization angle \\
14 & Uncertainty on the NVSS polarization angle \\
15 & NVSS catalog fitted deconvolved major axis \\
16 & Upper limit flag on the deconvolved major axis \\
17 & NVSS catalog fitted deconvolved minor axis \\
18 & Upper limit flag on the deconvolved minor axis \\
19 & S-PASS peak intensity, $I_\mathrm{SP}$ \\
20 & Uncertainty on the $I_\mathrm{SP}$ \\
21 & S-PASS polarized intensity  \\
22 & Uncertainty on the S-PASS polarized intensity \\
23 & S-PASS fractional polarization, $\pi_\mathrm{SP}$ \\
24 & Uncertainty on the $\pi_\mathrm{SP}$ \\
25 & Upper limit flag on the $\pi_\mathrm{SP}$ \\
26 & S-PASS polarization angle \\
27 & Uncertainty on the S-PASS polarization angle \\
28 & Spectral index derived from NVSS and S-PASS \\
29 & Uncertainty on the spectral index \\
30 & Depolarization, $D$ \\
31 & Uncertainty on the $D$ \\
32 & Taylor et al. 2009 rotation measure, $\textrm{RM}_\mathrm{T}$ \\
33 & Uncertainty on the $\textrm{RM}_\mathrm{T}$ multiplied by 1.22 \\
34 & The NVSS \& S-PASS rotation measure, $\textrm{RM}_\mathrm{NS}$ \\
35 &  Uncertainty on the $\textrm{RM}_\mathrm{NS}$  \\
36 &  Rotation measure difference, $\Delta RM$ \\
37 & Uncertainty on the $\Delta RM$ \\
38 & Median $\textrm{RM}_\mathrm{T}$  \\
39 & Number of sources contributed to the median $\textrm{RM}_\mathrm{T}$ \\
40 & Redshift from Hammond et al. 2012 \\
41 & WISE catalog W1 (3.4 micron) magnitude \\
42 & Uncertainty on the W1 \\
43 & W1 detection signal to noise ratio \\
44 & WISE catalog W2 (4.6 micron) magnitude \\
45 & Uncertainty on the W2 \\
46 & W2 detection signal to noise ratio \\
47 & WISE catalog W3 (12 micron) magnitude \\
48 &  Uncertainty on the W3 \\
49 & W3 detection signal to noise ratio \\
\end{deluxetable}

\subsection{Rotation measures}\label{rm0}
\begin{figure}[h]
\centering
\includegraphics[scale=0.51]{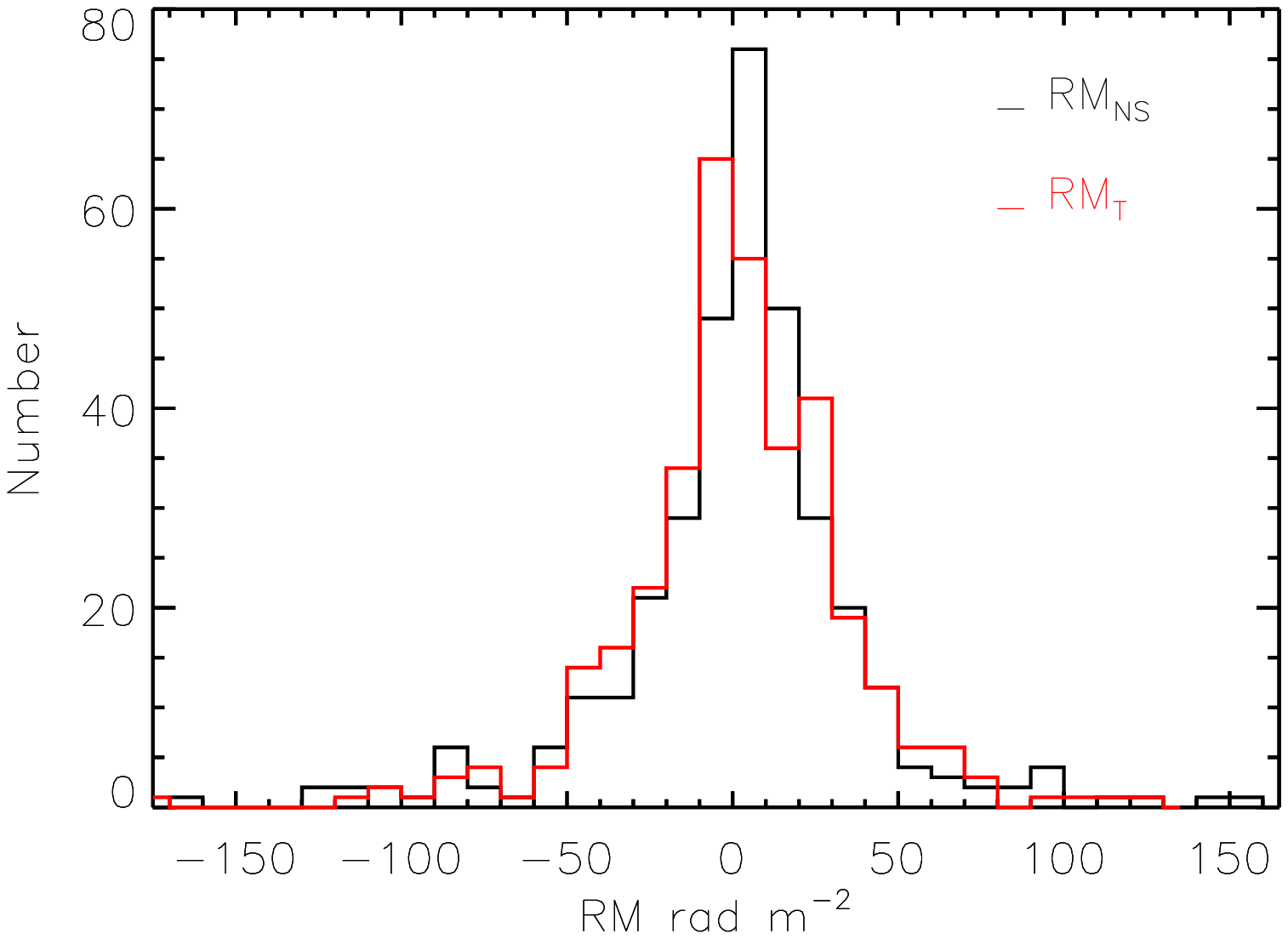}
\includegraphics[scale=0.51]{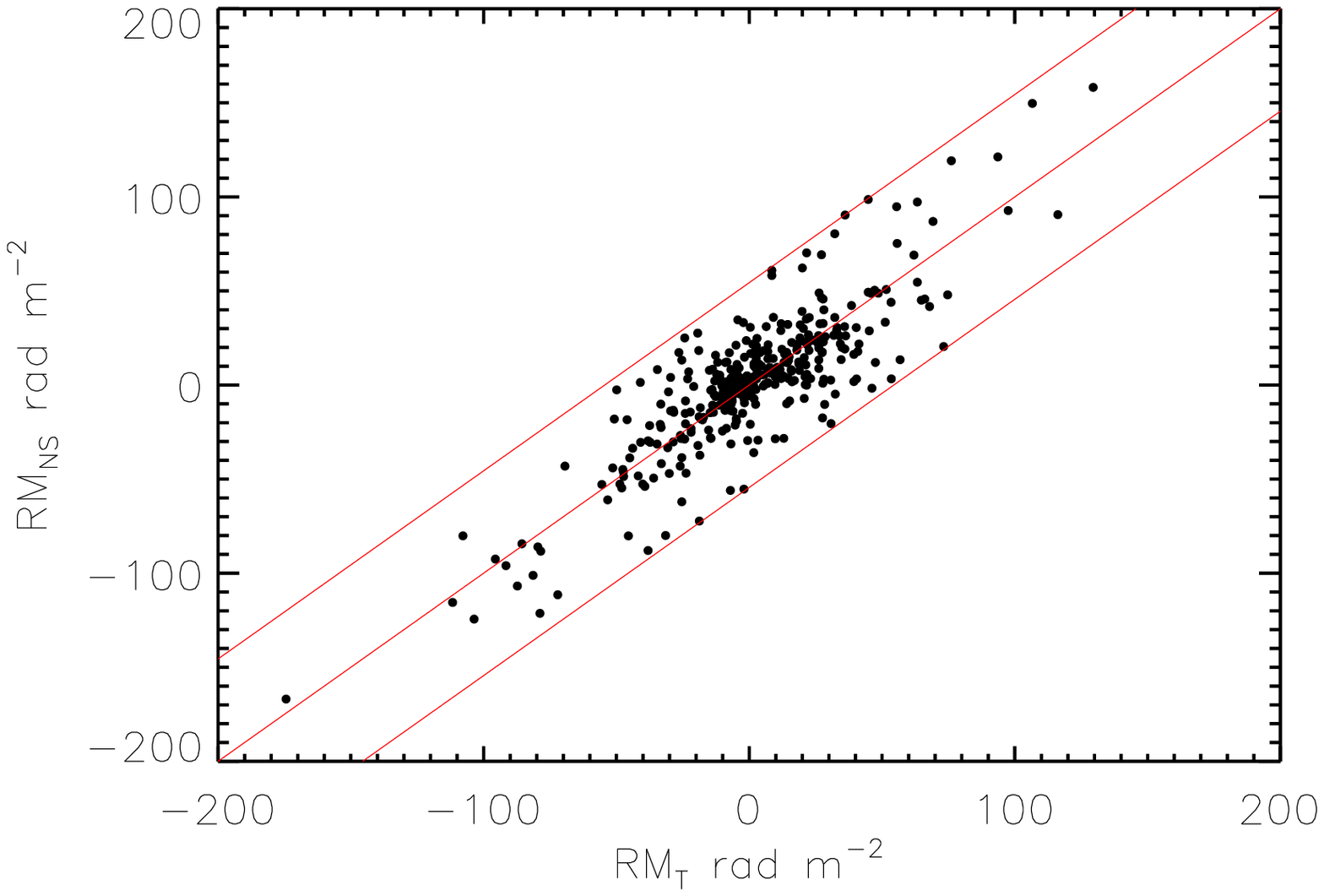}
\caption{Top: The distributions (top) and the scatter diagram (bottom) of the NVSS, S-PASS rotation measures, RM$_\mathrm{NS}$ versus TSS09 RM$_\mathrm{T}$ for the 364 common objects. The three red solid lines in the bottom show one-to-one relations for the three cases of $n= [-1,0,1]$.}
\label{fig:rmhist}
\end{figure}

The distribution of  $\textrm{RM}_\mathrm{NS}$ calculated based on NVSS and S-PASS (black) and the Taylor et al. rotation measures,  $\textrm{RM}_\mathrm{T}$, (red) for the same objects are shown in Figure \ref{fig:rmhist}. Both distributions are very similar in shape. Their medians are $3.6\pm 2.0$  and $0.5 \pm 1.9 $ rad m$^{-2}$ , respectively, while their standard deviations are 38.4 and 36.4 rad m$^{-2}$. Some of the scatter in the $\textrm{RM}$ distributions could be due to the uncertainty of the measurements.  \citep{2011ApJ...726....4S}. However, the median error on the  $\textrm{RM}_\mathrm{T}$ for the small bright sample of 364 objects in this work is only $\sigma_\mathrm{T}=3.5$ rad m$^{-2}$. The median measurement uncertainty estimated for  $\textrm{RM}_\mathrm{NS}$ is even smaller, $\sigma_\mathrm{NS}=1.6$ rad m$^{-2}$. Subtracting the median errors from the observed standard deviation of the $\textrm{RM}$ distributions in quadrature result in residual standard deviations of $36.2$ rad m$^{-2}$ and $38.36$ rad m$^{-2}$ for  $\textrm{RM}_\mathrm{T}$ and  $\textrm{RM}_\mathrm{NS}$ respectively, and largely represent the spread in Galactic foregrounds.
     
\subsection{Distribution of fractional polarization and depolarization}\label{D_distribution}
\begin{figure}[h]
\centering
\includegraphics[scale=0.51]{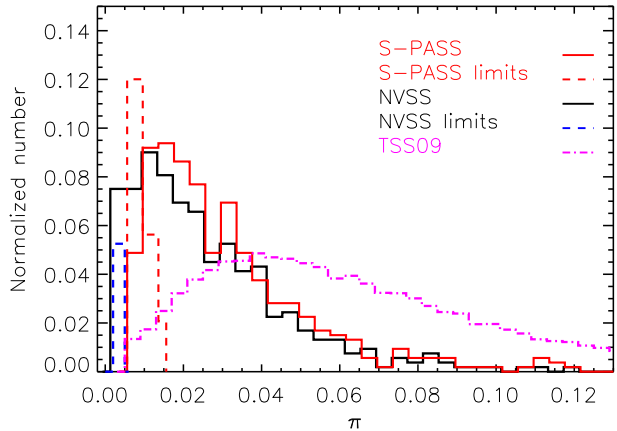}
\caption{Normalized histograms of fractional polarizations, $\pi$, for 416 objects with detected polarization in both NVSS and S-PASS and the upper limits. The black and red solid lines represent the NVSS and S-PASS distributions of objects with detected polarizations while the dashed blue and red lines sketches the distribution of upper limits of NVSS and S-PASS polarizations. For comparison we also show the NVSS fractional polarization distribution of the TSS09 catalog 37543 sources with dotted-dashed magenta line.  \label{fig:fphist}}
\end{figure}

The median NVSS (S-PASS) fractional polarization of all 533 objects is $\bar{\pi}=0.017$ $(0.020)$ including the upper limits. There are $505$ ($428$) objects with detected NVSS (S-PASS) polarization ($P > 3\sigma_p$ and $\pi > \epsilon $).  However, 416 of these objects are detected in both NVSS and S-PASS. The distributions of NVSS and S-PASS fractional polarization of these 416 objects are shown in Figure \ref{fig:fphist}. The median (and standard deviation) of NVSS and S-PASS fractional polarization of these common objects are 0.022 (.022) and 0.025 (0.023), respectively.  
 Although the median values of $\pi_\mathrm{SP}$ and $\pi_\mathrm{NV}$ are very close, the median value of their ratio (the median depolarization) is not necessarily equal to one. 

The TSS09 catalog was limited to sources with sufficient signal:noise in polarization, and is thus biased towards much higher fractional polarizations (median $\bar{\pi}_T \sim 0.06$) than our catalog, which is $\sim$3.5 times lower, including both measurements and upper limits.    

\begin{figure}[h]
\centering
\includegraphics[scale=0.51]{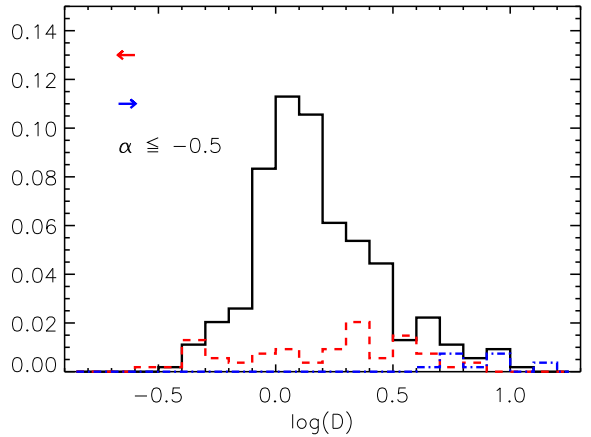}
\includegraphics[scale=0.51]{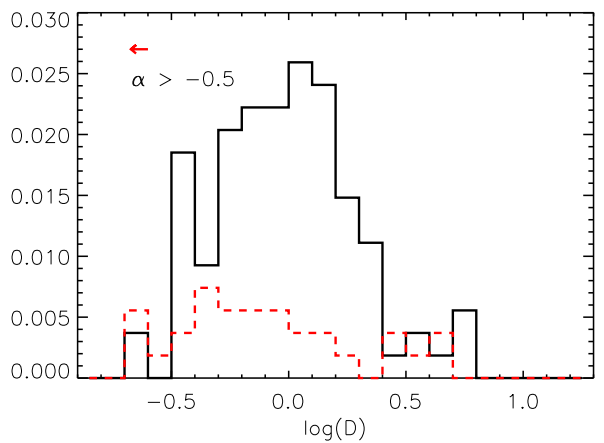}
\caption{Distributions of $\log(D)$ normalized to the total number of objects for steep (top) and flat (bottom) spectrum sources. Black solid histogram represents objects with detected polarization in both NVSS and S-PASS. The red histogram with dashed line is the distribution of the upper limits in depolarization. The lower limits are shown with dotted-dashed blue line. The two red and blue arrows show the direction of movement for the upper and lower limits. }
\label{fig:depol}
\end{figure}

Figure \ref{fig:depol} shows the normalized distribution of $\log(D)$ for steep and flat spectrum sources separately. Objects with both S-PASS and NVSS detected polarizations are shown in solid black,  and have median depolarizations of $\bar{D}=1.4$ and $\bar{D}=0.9$ for 315 steep and 101 flat sources respectively. The depolarization distribution of steep spectrum sources is skewed toward large values of $D$. Almost 28\% of steep spectrum (24\% of all) objects have $D \ge 2$, and only 2\% have $D \le 0.5$. On the other hand, flat spectrum sources include both depolarized and re-polarized objects. There are 17\% and 13\% of flat spectrum sources with  $D \ge 2$ and  $D \le 0.5$ respectively. The results of the statistical tests presented in Table \ref{table2} confirm that steep and flat spectrum sources do not have the same depolarization distributions. 

The red dashed histogram in Figure \ref{fig:depol} shows the normalized distribution of 58 steep spectrum and 31 flat spectrum objects with upper limits on the depolarization. These sources have S-PASS polarizations less than $3\sigma$ or $\pi_\mathrm{SP} < \epsilon_\mathrm{SP}$ but are detected in NVSS polarization.  
The 12 steep spectrum objects with NVSS $P < 3\sigma_p$ or $\pi_\mathrm{NV} < \epsilon_\mathrm{NV}$ and detected S-PASS polarization are treated as lower limits on the depolarization. The dotted dashed blue line show the distribution of the lower limits in Figure \ref{fig:depol}. In total, 16 objects are detected in neither NVSS nor in S-PASS polarizations and we do not show them in  Figure \ref{fig:depol}. 

\begin{figure}[h]
\centering
\includegraphics[scale=0.51]{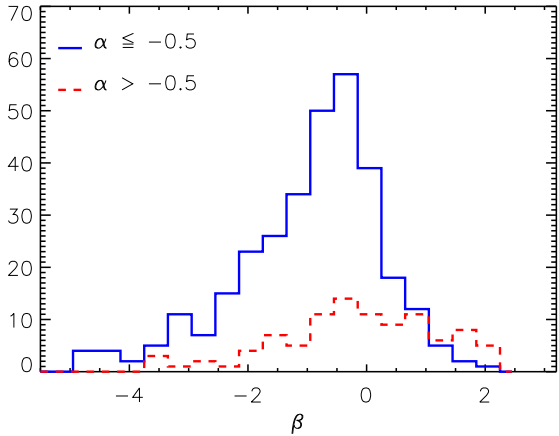}
\caption{Distribution of the polarization spectral index $\beta$ as introduced in \cite{2014ApJS..212...15F} assuming a power law depolarization model. The solid blue and dashed red lines represent the steep and flat spectrum sources.  
\label{beta}}
\end{figure}

\cite{2014ApJS..212...15F} used their multi wavelength polarization spectra and derived an equivalent power law polarization spectral index $\beta$, where $\pi \propto \lambda^{\beta}$.  As long as the power law model is assumed our depolarization parameter $D$ and $\beta$ are related such that $\log(D)=\log(\frac{\lambda_\mathrm{SP}}{\lambda_\mathrm{NV}}) \beta$ where the $\lambda_\mathrm{SP}$ and $\lambda_\mathrm{NV}$ are the average wavelengths of the S-PASS and NVSS surveys respectively.  \cite{2014ApJS..212...15F} found weak evidence of a bimodal distribution for $\beta$ of steep spectrum objects. We do not see any sign of bimodal depolarization within objects with $\alpha < -0.5$, as shown in Figure \ref{beta}. The $\beta$ distribution of steep spectrum objects is single-peaked but asymmetric with a longer tail toward depolarized objects.  
As will be discussed later, the majority of steep spectrum sources in our sample can be classified as IR AGNs according to their infrared colors. A more complete sample which also includes radio galaxies with infrared colors of normal ellipticals can confirm  if the weak bimodal depolarization observed by \cite{2014ApJS..212...15F}  is real.   

\begin{figure}[h]
\centering
\includegraphics[scale=0.51]{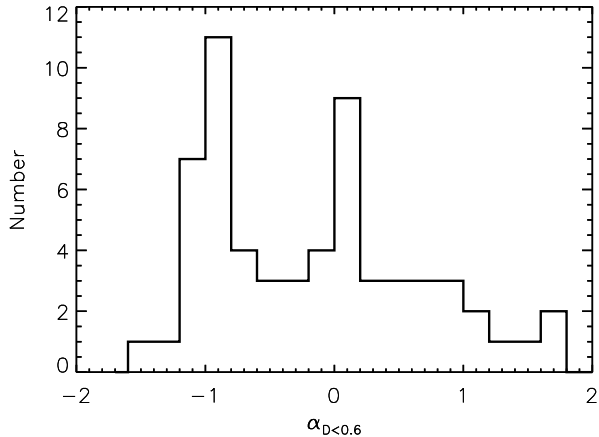}
\caption{ Spectral index distribution of the re-polarized objects, $D< 0.6$, including 24 sources with detection in $\pi_\mathrm{NV}$ but only upper limits on $\pi_\mathrm{SP}$. While it seems there are two separate populations of re-polarized sources with flat and steep spectrums, the majority of them, 61\%, have $\alpha \ge -0.5$.  
\label{sphstrep}}
\end{figure}

\begin{figure}[h]
\centering
\includegraphics[scale=0.51]{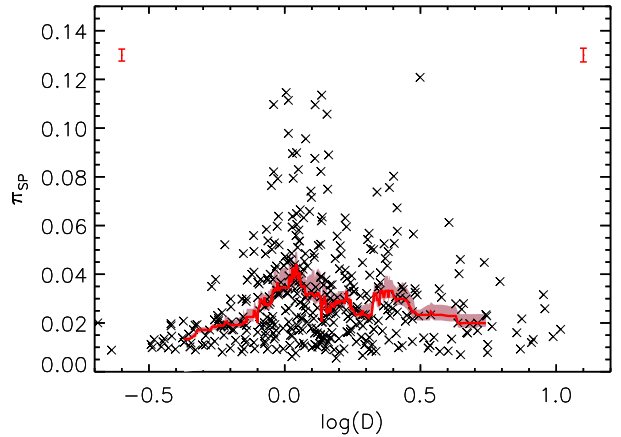}
\caption{S-PASS fractional polarization versus $\log(D)$. The red solid line represents the running median of $\pi_\mathrm{SP}$  calculated in bins of $N=30$ objects in $\log(D)$ space and the dark-pink shaded region is the estimated uncertainty on the running medians calculated as $|M-[p16, p84]|/\sqrt(N)$ where $M$ is the median value and $[p16, p84]$ are the 16 and 84 percentiles. The error bars on the left and right upper corners are the medians of the intrinsic uncertainties in $\pi_\mathrm{SP}$ for the two half of data in $\log(D)$. 
}
\label{fig:depfp}
\end{figure}

We also looked at the combined sample of steep and flat spectrum sources and classified them into three depolarization categories. The choice of the depolarization boundaries is somewhat arbitrary. However, we designed the three depolarization categories to isolate the peak observed in Figure \ref{fig:depfp}, as is discussed below. Sources with $ 0.6 < D < 1.7 $ have median spectral index of $\bar{\alpha} \sim -0.82$ while sources with $D \ge 1.7$ shows a slightly steeper median spectrum with $\bar{\alpha} \sim -0.9$. The spectral slope is mostly flat for re-polarized objects with $D \le 0.6$, with a median $\bar{\alpha} \sim -0.1$. However, there are 14 re-polarized objects with steep spectral indices, $\alpha < -0.5$. This is consistent with \cite{2014ApJS..212...15F} who also found a small population of steep spectrum re-polarized sources. Figure \ref{sphstrep} shows the distribution of the spectral indices of re-polarized objects. We also included 24 objects with detection in $\pi_\mathrm{NV}$ but only upper limits on $\pi_\mathrm{SP}$. Figure \ref{sphstrep} suggests there are two separate populations of re-polarized sources with flat and steep spectra. Including the mentioned upper limits on $D$, 61\% of re-polarized sources have $\alpha \ge -0.5$ (i.e., flat). 

To understand the relation between fractional polarization and depolarization, we plotted $\pi_\mathrm{SP}$ versus $\log(D)$, and calculated the running medians in bins of  30 objects (Figure \ref{fig:depfp}). There is an apparent peak for S-PASS fractional polarization at $log(D )\sim 0$, while both depolarized and re-polarized sources show weaker $\pi_\mathrm{SP}$ than sources with fractional polarization above 6\%. Both KS and Spearman rank coefficient tests on the $|\log(D)|$ and $\pi_\mathrm{SP}$  confirm this anti-correlation. We also used two subsamples with $\log(D) >0$ and $\log(D) <0$ and performed the two KS and Spearman rank tests on each subsample separately. The results confirmed that fractional polarizations are higher in the vicinity of $\log(D)\sim 0$ in each subsample. However, the correlation between $|\log(D)|$ and $\pi_\mathrm{SP}$ of the subsample with $\log(D) >0$ became uncertain when only including the contamination clean sample of the robustness test. Table \ref{table2} summarizes the results of these statistical tests.

\begin{figure}[h]
\centering
\includegraphics[scale=0.51]{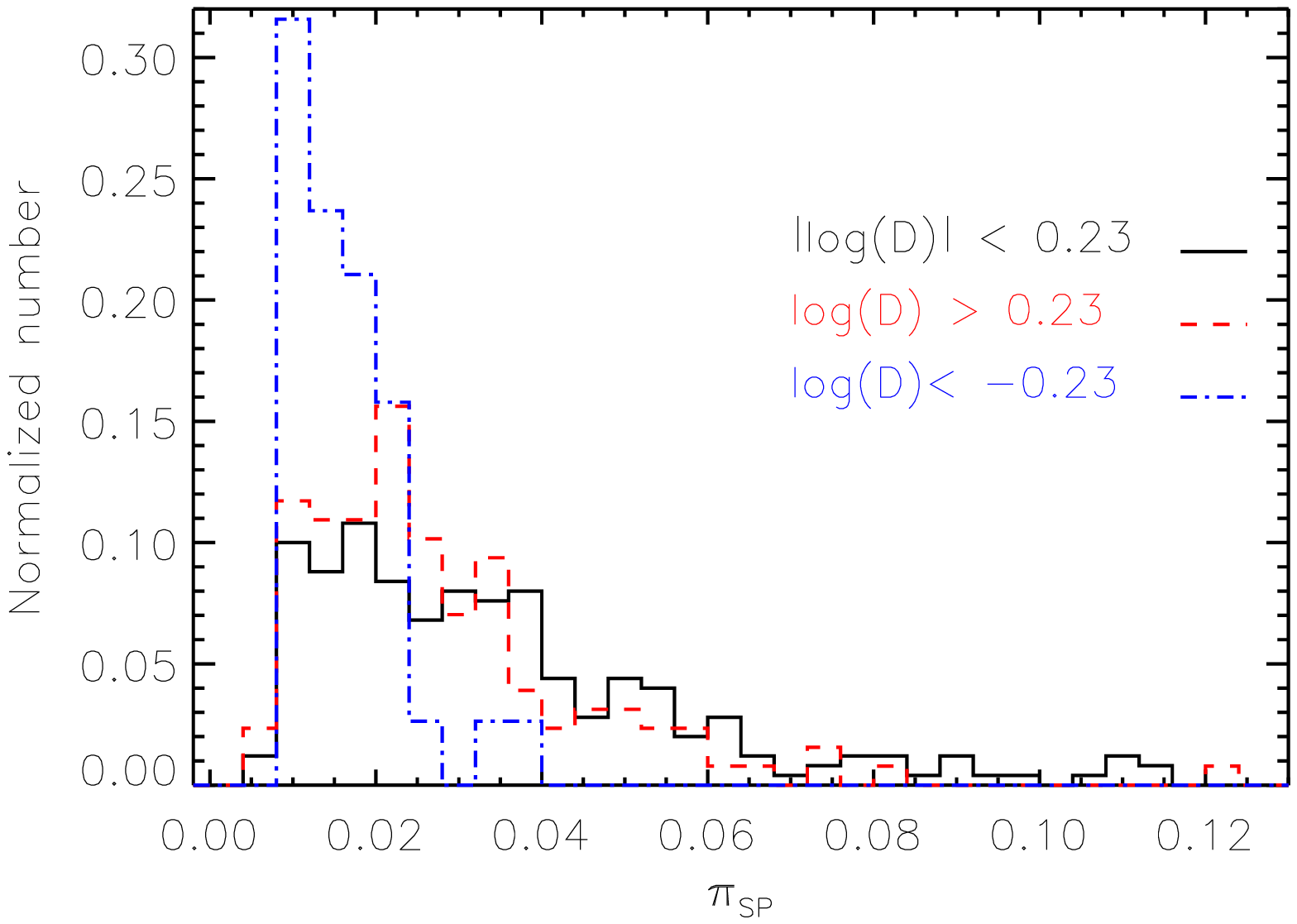}
\includegraphics[scale=0.51]{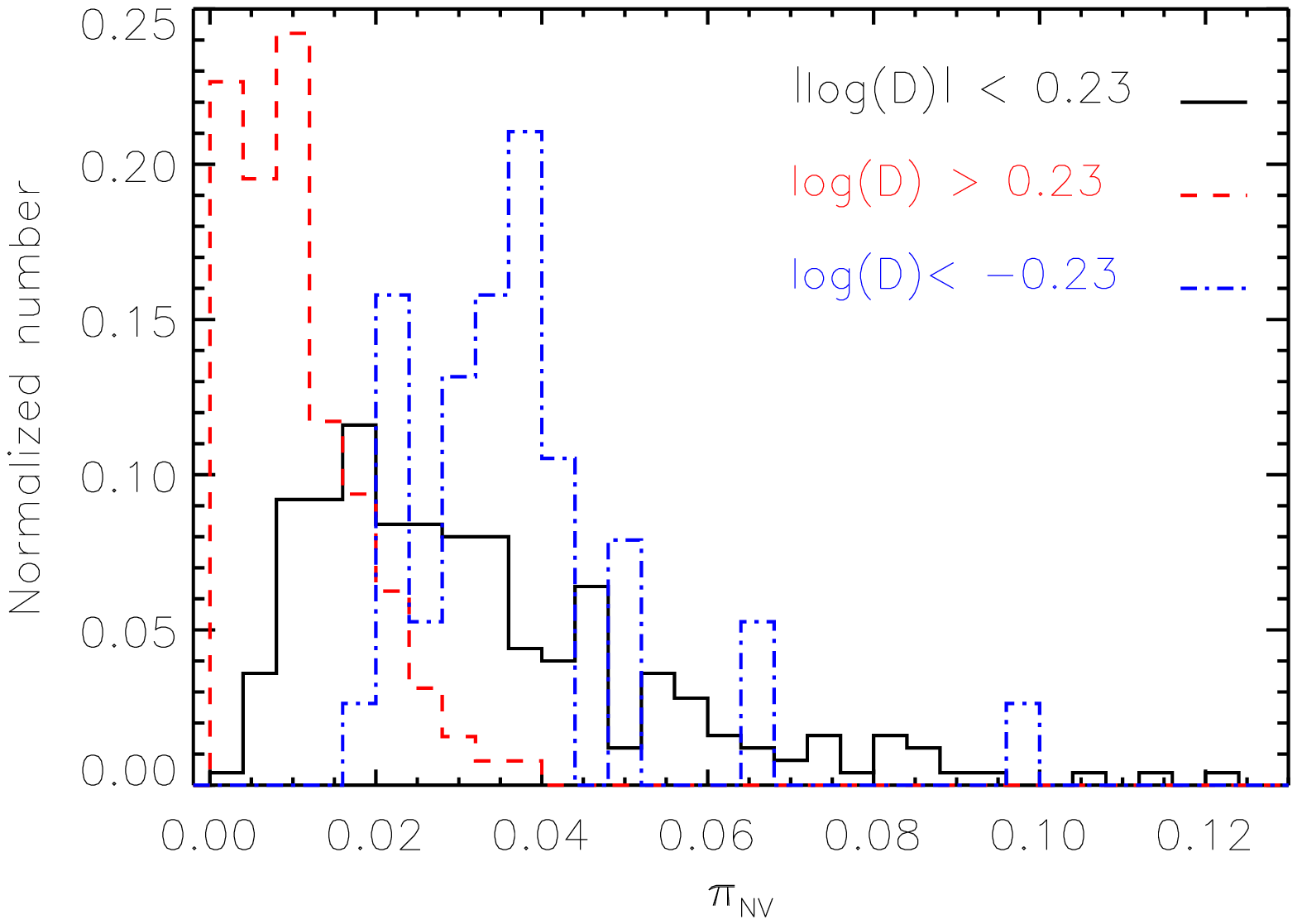}
\caption{Normalized S-PASS (upper) and NVSS (lower) fractional polarization distribution for objects with $|\log(D) < 0.23|$ (solid black), $\log(D) > 0.23$ (dashed red) and $\log(D)<-0.23$ (dotted-dashed blue).
\label{fig:fphist23}}
\end{figure}

Figure \ref{fig:fphist23} shows the S-PASS (top) and NVSS (bottom) fractional polarization distributions for three sub-samples with $|\log(D)| \le 0.23$, $\log(D) > 0.23$ and $\log(D)<-0.23$. Objects with $\log(D) \sim 0$ have almost the same distribution in both S-PASS and NVSS (by construction) with median fractional polarizations of $\bar{\pi}_\mathrm{SP}=0.030$ and $\bar{\pi}_\mathrm{NV}=0.028$ while depolarized sources have smaller medians, $\bar{\pi}_\mathrm{SP}=0.024$ and $\bar{\pi}_\mathrm{NV}=0.009$ with an offset between NVSS and S-PASS as expected. 

Objects with re-polarization show more complicated behavior. They have a median $\bar{\pi}_\mathrm{SP}=0.015$ and $\bar{\pi}_\mathrm{NV}=0.035$. By definition the median degree of polarization of a sample of re-polarized sources is expected to be higher at 1.4 GHz than 2.3 GHz. 
It is possible that the true median $\bar \pi_\textrm{SP}$ and $\bar \pi_\textrm{NV}$ are lower than the above values because we would have systematically excluded re-polarized  objects with $\pi_\mathrm{SP}$ less than the detection limit. This results in over estimating the median fractional polarization of re-polarized sources in both NVSS and S-PASS.

\subsection{Total intensity and fractional polarization}\label{ip}
\begin{figure}[h]
\centering
\includegraphics[scale=0.51]{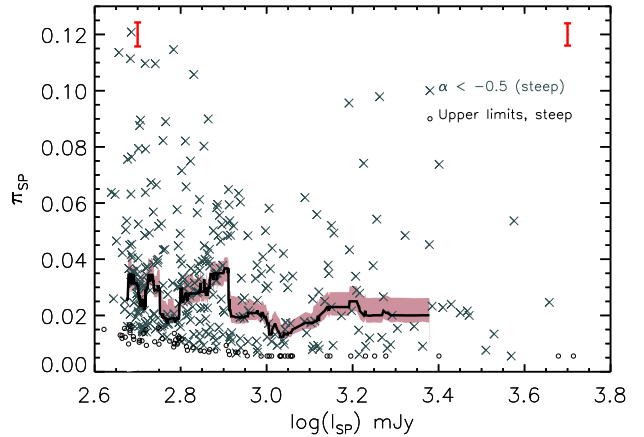}
\caption{S-PASS fractional polarization of only steep spectrum($\alpha < -0.5$) versus their total intensity.  The open circles represent the upper limits on the degree of polarization. The black solid line is the running medians of $\pi_\mathrm{SP}$ including the upper limits and the dark-pink shaded region is the estimated uncertainty on the running medians. The red error bars in upper right and left corners show the median intrinsic uncertainties of $\pi_\mathrm{SP}$ for the two half of the data in $\log(I)$ space. 
\label{fig:Ifp}}
\end{figure}

Our sample includes total intensities from 0.42 to 10 Jy, which gives us the opportunity to study possible correlations between the fractional polarization and total intensity. 
As listed in Table \ref{table2} both KS and Spearman tests suggest there is a weak anti-correlation between $\pi_\mathrm{SP}$ and $I_\mathrm{SP}$ of the whole sample of sources at 2.3 GHz. More investigation revealed that is true for  steep spectrum ($\alpha < -0.5$) sources alone, while it disappears for flat spectrum ($\alpha \ge -0.5$) objects. The anti-correlation among steep spectrum sources became weaker and more uncertain when only including the contamination clean sample of the robustness test, and thus should be treated as a suggestive trend only. Figure \ref{fig:Ifp} shows the S-PASS $\pi_\mathrm{SP}$ of only steep spectrum sources versus their logarithm of total intensity. The calculated running medians (including the upper limits on $\pi_\mathrm{SP}$ to avoid any selection bias due to our total intensity cut) are shown as well. Objects with $\alpha < -0.5$ and $\log(I_\mathrm{SP}) < 2.9$ have median of $\bar{\pi}_\mathrm{SP} \sim 0.03$ while sources with larger total intensity are less polarized with medians of $\bar{\pi}_\mathrm{SP} \sim 0.02$.  

To shed light on a possible physical origin of the observed anti-correlation we calculated the luminosities, based on the 261 objects in our sample which have redshifts in the  \cite{2012arXiv1209.1438H} catalog.  222 of these sources are detected in both NVSS and S-PASS polarization. 
Using our spectral indices, we calculated the K-corrected 2.3 GHz luminosities. The 141 steep spectrum objects have median luminosity of $L_{steep}=1.7\times 10^{27}$ WHz$^{-1}$.  Although there is a nominal difference between $\bar{\pi}_\mathrm{SP}$ for  higher and lower luminosities (2.6\% and 2.2\%, respectively), these do not appear statistically significant.  There is also no statistically significant difference in $|\log(D)|$ for the high and low luminosity steep spectrum sources.  

The 81 flat spectrum sources are at higher redshifts, on average, and have a median luminosity of $\bar L_{flat}=3.0 \times 10^{27}$ W Hz$^{-1}$. 

\subsection{Correlation between  $\textrm{RRM}$, $\Delta \textrm{RM}$, $\pi$ and $D$}\label{deltarm}
There are two measures to characterize the Faraday effects that are either local to the  source or in the intervening IGM medium, the residual rotation measure  $\textrm{RRM}$, which takes out the Galactic foreground contribution to the observed  $\textrm{RM}$, and  $\Delta \textrm{RM} \equiv \textrm{RM}_\mathrm{T}-\textrm{RM}_\mathrm{NS}$, which sheds light on the frequency dependency of the $\textrm{RM}$. The absolute value of $|\Delta RM|$ is an indicator of the Faraday complexity of the source and its environment. As explained in the following, we found that $\Delta \textrm{RM}$ is anti-correlated with $\pi$ and correlated with $|\log(D)|$.

\begin{figure}[h]
\centering
\includegraphics[scale=0.51]{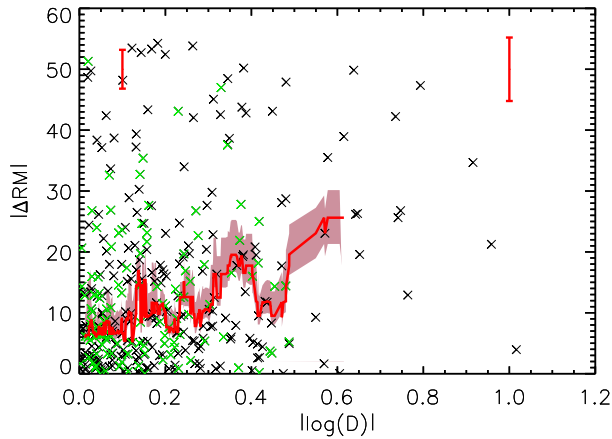}
\caption{Absolute difference between rotation measures calculated in this work and in TSS09, $|\Delta \textrm{RM}|$ versus $|\log(D)|$. Black and green crosses represent depolarized and re-polarized objects respectively.
The solid red line is the running median of $|\Delta \textrm{RM}|$ calculated for bins of 23 objects in $|\log(D)|$ space which include both depolarized and re-polarized sources and the dark-pink shaded region is the estimated uncertainty on the running medians. The error bars on the left and right upper corners are the medians of intrinsic uncertainties in $|\Delta \textrm{RM}|$ for the two halves of the data. 
\label{fig:drm2new}}
\vskip 1mm
\end{figure}

Faraday complex sources, i.e, those with multiple $\textrm{RM}$ components should be both depolarized and have polarization angles which may not vary linearly with $\lambda^2$. We therefore examined the possible correlation between depolarization and $|\Delta \textrm{RM}|$. Figure \ref{fig:drm2new} shows $|\Delta \textrm{RM}|$ versus $|\log(D)|$ for all objects with detected polarization in both NVSS and S-PASS. The running medians of the $|\Delta \textrm{RM}|$ calculated in bins of $|\log(D)|$ show an evolution.  To quantify this, 
we calculated the Spearman rank, which yielded a correlation coefficient of $r_s=0.23$ and p-value of $p = 0.00003$ establishing that depolarization and non-$\lambda^2$ polarization angle behavior are related.  

\begin{figure}[h]
\centering
\includegraphics[scale=0.51]{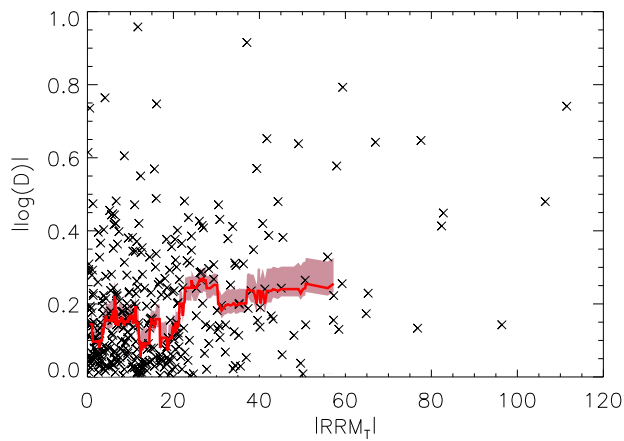}
\caption{ The absolute residual rotation measures, $|\textrm{RRM}_\mathrm{T}|$ versus the $|\log(D)|$. The red solid line which represent the running medians of $|\log(D)|$, shows an increase with raising $|\textrm{RRM}_\mathrm{T}|$. The dark-pink shaded region is the estimated uncertainty on the running medians.\label{rrmdep}}
\end{figure}

A large $\textrm{RM}$ beyond the Galactic foreground $\textrm{RM}$ screen could also indicate the presence of Faraday complexity and depolarization.  To estimate this, we removed the Galactic contribution by subtracting the median $\bar{\textrm{RM}}$ within 3 degrees of each target (excluding the target itself), using the TSS09 catalog.  This yields the residual rotation measure,  $\textrm{RRM}_\mathrm{T} \equiv \textrm{RM} -\bar{\textrm{RM} }$. Subtracting the median $\bar{\textrm{RM}}$ is not the best method to estimate the extragalactic component of the $\textrm{RM}$ as discussed in \cite{2015A&A...575A.118O}. However, for objects above the Galactic latitude of $|b| > 20$ degrees, which is true for most of our sample, the difference between the \cite{2015A&A...575A.118O} recipe and our method is small. As shown in Figure \ref{rrmdep}, we find the Spearman rank coefficient of $r_s=0.21$ and p-value of $7\times10^{-5}$ which suggests a correlation between $|\textrm{RRM}_\mathrm{T}|$ and $|$log(D)$|$.  However, our robustness test on the clean sample failed to confirm such a trend. Thus, only $|\Delta \textrm{RM}|$ shows a clear sign of a correlation with depolarization.

\begin{figure}[h]
\centering
\includegraphics[scale=0.51]{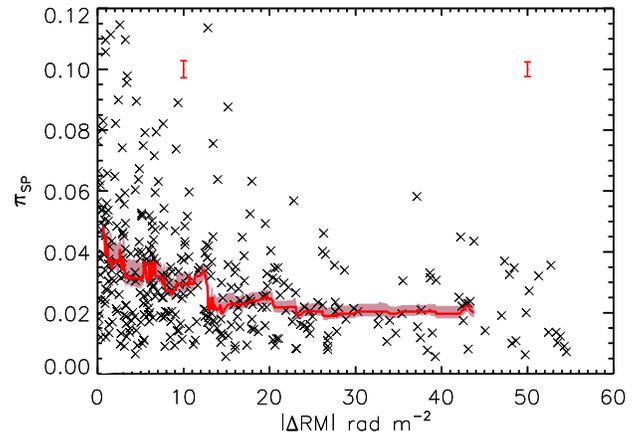}
\caption{ S-PASS fractional polarization versus the $|\Delta \textrm{RM} |$ which is a representation of the Faraday structure. \label{drmp}}
\end{figure}

We also found, the 1.4 GHz and 2.3 GHz fractional polarizations show moderate anti-correlations with $|\Delta \textrm{RM} |$, as shown in Figure \ref{drmp} and listed in Table \ref{table2}. Thus, depolarization does reduce the fractional polarizations at these frequencies,  although the dominant role of field disorder is discussed in Section \ref{obs}.
Moreover, the Spearman rank test with $r_s=-0.25$ and p-value of $<10^{-5}$ suggest an anti-correlation between $|\textrm{RRM}_\mathrm{T}|$ and $\pi_\textrm{NV}$. However, our robustness test failed to confirm this significance.

\subsection{Polarization, depolarization and the object angular extent}\label{extent}
\begin{figure}[h]
\centering
\includegraphics[scale=0.51]{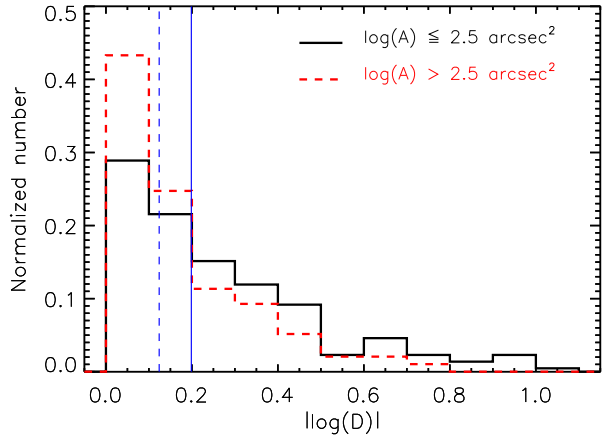} 
\caption{The $|\log(D)|$  distributions of the unresolved (black solid) and extended (dashed red) steep spectrum objects in the NVSS survey. The de-convolved surface area thresholds $\log(A) \le 2.5$ arcsec$^2$ and $\log(A) > 2.5$ arcsec$^2$ are used to separate unresolved and extended sources, and the two vertical blue solid and dashed lines represent the medians of $|\log(D)|$  for the two samples respectively.
\label{fig:dsize}}
\end{figure}

To study how the morphology of a system affects the depolarization, we used total intensity deconvolved areas ($A$) derived from the NVSS catalog \cite{1998AJ....115.1693C}. Flat spectrum objects in our sample are unresolved in the NVSS synthesized beam while steep spectrum objects include both resolved and unresolved sources. For the steep spectrum sources, Figure \ref{fig:dsize} shows the distributions of the absolute $|\log(D)|$ for two sub-samples -  unresolved and resolved sources with the dividing line at $\log(A) = 2.5$ arcsec$^2$.  On average, resolved sources have smaller $|\log(D)|$ with median of $0.12$ compared to $0.20$ for unresolved sources. The scatter of the two samples is almost the same with standard deviation of $0.21$. 
Beam depolarization should only play a small role, because most resolved sources are only slightly resolved.  

\begin{figure}[h]
\centering
\includegraphics[scale=0.51]{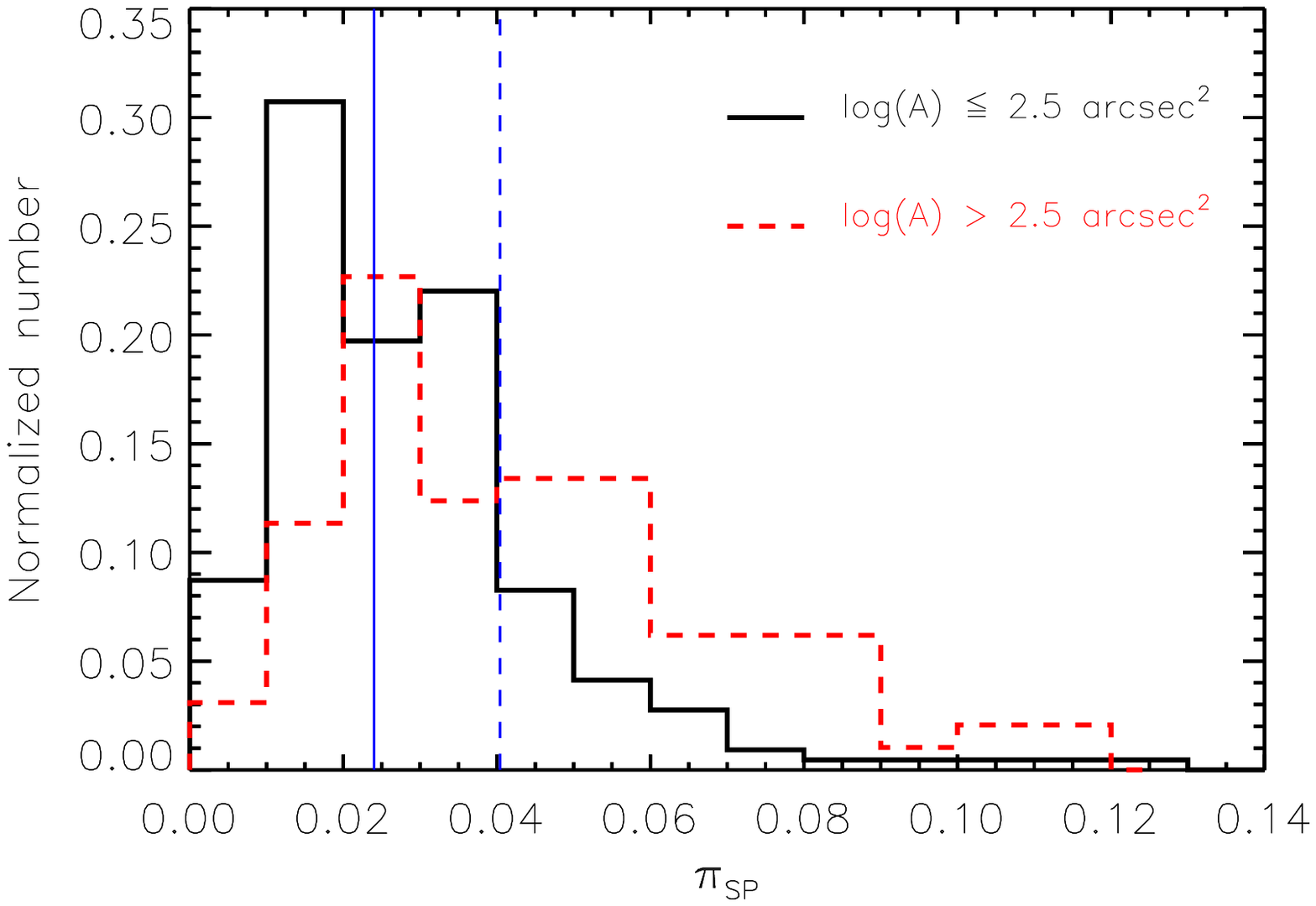}
\caption{The $\pi_\mathrm{SP}$ distributions of the unresolved (black solid) and extended (dashed red) steep spectrum objects in the NVSS survey. The de-convolved surface area thresholds $\log(A) \le 2.5$ arcsec$^2$ and $\log(A) > 2.5$ arcsec$^2$ are used to separate unresolved and extended sources, and the two vertical blue solid and dashed lines represent the medians of $\pi_\mathrm{SP}$ for the two samples respectively.
\label{fig:sizefp}}
\end{figure}

We also looked at the dependence of fractional polarization on size. Figure \ref{fig:sizefp} shows the distributions of the S-PASS fractional polarization for the unresolved and extended samples of steep spectrum objects.  
On average, resolved and extended steep spectrum objects have 2.3 GHz fractional polarizations, $\bar \pi_\mathrm{SP} \sim 4\%$, two times larger than their unresolved counterparts.  Both KS and Spearman tests confirm a strong  strong positive correlation between $A$ and $\pi_\mathrm{SP}$ of steep spectrum objects.

\subsection{Spatial distribution of depolarization in the sky}\label{galD}
\begin{figure}
\centering
\includegraphics[scale=0.51]{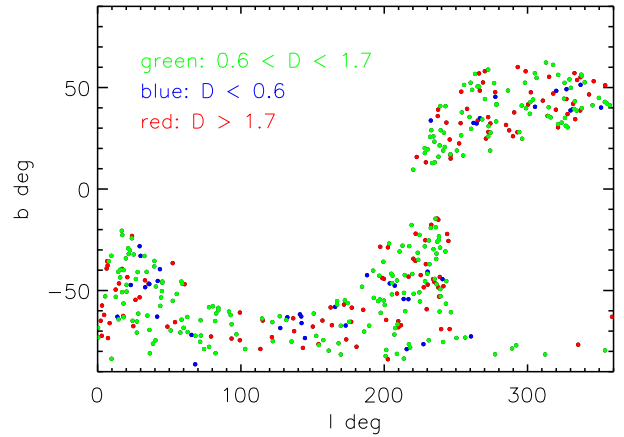}
\caption{Distribution of the 533 objects in the sky, color coded with the depolarization. $l$ and $b$ are the Galactic longitude and latitude coordinates in degrees. Black dots are objects that are not detected in either NVSS or S-PASS polarization. Green, red and blue  triangles are objects with depolarization $0.5< D <2$, $D > 2$ and $D < 0.5$ respectively.
\label{fig:galactic}}
\end{figure}

We  carried out a brief investigation to see if the depolarization properties in our sample were related to their position in Galactic coordinates. Figure \ref{fig:galactic} shows the distribution of 533 objects in the sky, color coded with respect to their depolarization. Visual inspection does not reveal any obvious over-density of depolarized or re-polarized objects. 

We also calculated the auto correlation between depolarization and angular separation, and the two point angular correlation function for the most depolarized and least depolarized sources.  None of these showed any evidence for clustering of depolarization in space. Similarly, the two point angular correlation functions for the highest and lowest fractional polarizations at 2.3~GHz revealed no clustering. Other work has identified some positional dependence to polarizations in the NVSS catalog.  \cite{2007ApJ...663L..21S}  discovered regions with angular scales of $\sim 10$ degrees in which the density of the polarized sources drops by a factor of 2-4. They named these regions the ``polarization shadows,'' and found that some of them are associated with the Galactic HII regions while the rest are related to the depolarized areas in the diffuse Galactic radio emission. All polarization shadows in \cite{2007ApJ...663L..21S}  are located within the Galactic plane at $|b| < 20 $ degrees except one which is at Galactic ($l=5$, $b=+24$). Almost all of the objects in our sample have Galactic latitudes of $|b| > 20 $ degrees, and none are located around ($l=5$, $b=+24$), so the Galactic polarization shadows likely do not affect the current work. However, it is interesting to search for high latitude Galactic diffuse emissions in smaller scales and their probable signature on the depolarization of the extragalactic sources in future surveys and larger samples with higher number density. 

\tabcolsep=0.04cm
\tabletypesize{\scriptsize}
\begin{deluxetable*}{ccccc|cc}[h]
\centering
\tablecolumns{7}
\tablewidth{180mm}
\tablecaption{Results of non-parametric statistical tests with simulated p-values.  \label{table2}}
\tablehead{
\colhead{{Parameters}} & \colhead{Constraint}&\colhead{KS distribution} & \colhead{KS samples} & \colhead{KS p-value} & \colhead{Spearman rank} & \colhead{p-value} \\
\colhead{} & & & \colhead{ } & \colhead{ simulated}  & \colhead{correlation coefficient} &\colhead{simulated} }
\startdata
*$\pi_\mathrm{SP}$ \& $\alpha$ &  -   & $\pi_\mathrm{SP}$ & $\alpha \lessgtr -0.5$ &$<0.00001 $ &  $-0.24$& $<0.00001$  \\[5pt]
*$\pi_\mathrm{SP}$ \& $Area$ & $\alpha <-0.5$ & $\pi_\mathrm{SP}$ & $A \lessgtr \bar A$ & $<0.00001$ & $0.36$  &  $<0.00001 $ \\[5pt]
*$\pi_\mathrm{SP}$ \& $|\log(D)|$ &   -   &  $|\log(D)|$ & $\pi_\mathrm{SP} \lessgtr \bar{\pi}_\mathrm{SP}$  & $<0.00001 $ & $-0.28$ & $<0.00001 $  \\[5pt]
?$|\log(D)|$ \& $\pi_\mathrm{SP}$ & $\alpha<-0.5$  & $|\log(D)|$ & $\pi_\mathrm{SP} \lessgtr \bar{\pi}_\mathrm{SP}$ &  $0.00050$& $-0.26$  & $ <0.00001$  \\[5pt]
$|\log(D)|$ \& $\pi_\mathrm{SP}$ &  $\alpha \ge -0.5$  & $|\log(D)|$ & $\pi_\mathrm{SP} \lessgtr \bar{\pi}_\mathrm{SP}$& $0.022$&$-0.37$  & $0.00019 $  \\[5pt]
?$D$ \& $\pi_\mathrm{SP}$ &$D>1$ & $D$ &  $\pi_\mathrm{SP} \lessgtr \bar{\pi}_\mathrm{SP}$ &  $ 0.00004$ &$-0.25$ & $0.0095 $  \\[5pt]
*$D$ \& $\pi_\mathrm{SP}$ & $D<1$ &  $D$ &  $\pi_\mathrm{SP} \lessgtr \bar{\pi}_\mathrm{SP}$ &  $ 0.00005$ & $0.50$  &  $< 0.00001 $  \\[5pt]
?$I_\mathrm{SP}$ \& $\pi_\mathrm{SP}$  &  -   &  $\pi_\mathrm{SP}$ &  $I_\mathrm{SP} \lessgtr \bar{I}_\mathrm{SP} $ & $<0.00001 $ & $-0.25$  &  $<0.00001 $  \\[5pt]
?$I_\mathrm{SP} $ \& $\pi_\mathrm{SP}$ & $\alpha<-0.5$ &  $\pi_\mathrm{SP}$ &  $I_\mathrm{SP} \lessgtr \bar{I}_\mathrm{SP} $&  $<0.00001 $  & $-0.25$  &  $<0.00001 $  \\[5pt]
{$I_\mathrm{NV}$ \& $\pi_\mathrm{NV}$} &  -   &$\pi_\mathrm{NV}$ & $I_\mathrm{NV}  \lessgtr \bar{I}_\mathrm{NV}  $& $0.050 $ &-0.13  & $ 0.0021$  \\[5pt]
{$I_\mathrm{NV} $ \& $\pi_\mathrm{NV}$} &$\alpha<-0.5$ & $\pi_\mathrm{NV}$ & $I_\mathrm{NV}  \lessgtr \bar{I}_\mathrm{NV}  $& $0.094 $  &-0.16  & $0.0013 $  \\[5pt]
{$I_\mathrm{NV}  $ \& $\pi_\mathrm{NV}$} & $\alpha \ge -0.5$& $\pi_\mathrm{NV}$ & $I_\mathrm{NV}  \lessgtr \bar{I}_\mathrm{NV}  $& $ 0.96$  &-0.04  & $0.67 $  \\[5pt]
{$I_\mathrm{NV} $ \& $|\log(D)|$} &$\alpha<-0.5$ & $|\log(D)|$& $I_\mathrm{NV}  \lessgtr \bar{I}_\mathrm{NV}  $& $ 0.33$ &0.08  & $0.18 $  \\[5pt]
{$I_\mathrm{SP} $ \& $\pi_\mathrm{SP}$} &$\alpha \ge -0.5$ & $\pi_\mathrm{SP}$ & $I_\mathrm{SP} \lessgtr \bar{I}_\mathrm{SP} $& $0.26 $  &-0.06  & $ 0.47$  \\[5pt]
{$L_\mathrm{SP}$ \& $|\log(D)|$} & $\alpha<-0.5$ &$L_\mathrm{SP}$ & $|\log(D)| \lessgtr 0.13$ &$0.010 $ & 0.12& $ 0.16$    \\[5pt]
{$L_\mathrm{SP}$ \& $\pi_\mathrm{SP}$} & $\alpha<-0.5$ & $\pi_\mathrm{SP}$ &$L_\mathrm{SP} \lessgtr \bar{L}_\mathrm{SP}$ & $0.21 $ & -0.13& $ 0.11$    \\[5pt]
{$L_\mathrm{SP}$ \& $\pi_\mathrm{SP}$} & $\alpha \ge -0.5$ & $\pi_\mathrm{SP}$ &$L_\mathrm{SP} \lessgtr \bar{L}_\mathrm{SP}$ & $ 0.32 $ & 0.07& $ 0.53$    \\[5pt]
*$|\Delta \textrm{RM} |$ \& $|\log(D)|$  & -   & $|\log(D)|$ & $|\Delta \textrm{RM} | \lessgtr |\Delta \overline{\textrm{R}}\textrm{M}| $ &  $ 0.0010$ &  $0.23$ & $0.00003 $  \\[5pt]
*$|\Delta \textrm{RM} | $ \& $\pi_\mathrm{SP}$ &  -  &  $|\Delta \textrm{RM} | $ &  $\pi_\mathrm{SP} \lessgtr \bar{\pi}_\mathrm{SP}$ & $<0.00001 $ & $-0.40$  &  $ <0.00001$  \\[5pt]
*$D$ \& $|\Delta \textrm{RM} |$ &  $D>1$& $D$& $|\Delta \textrm{RM} | \lessgtr |\Delta \overline{\textrm{R}}\textrm{M}| $  & $ 0.017 $ &   $0.26$ &  $  0.00009 $  \\[5pt]
$D$ \& $|\Delta \textrm{RM}|$ & $D<1$ &$D$ &  $|\Delta \textrm{RM}| \lessgtr |\Delta \overline{\textrm{R}}\textrm{M}|  $  &  $ 0.12$ &   -0.14 & $  0.12$  \\[5pt]
$|\log(D)|$ \& $|\textrm{RRM}_\mathrm{T}| $ &  -  &$|\log(\mathrm{D})|$ & $|\textrm{RRM}_\mathrm{T}| \lessgtr |\textrm{R}\overline{\textrm{RM}}_\mathrm{T}|$&$ 0.019$ & $0.21$  &  $0.00007 $  \\[5pt]
$|\textrm{RRM}_\mathrm{T}| $ \& $\pi_\mathrm{NV}$ &  -  &$\pi_\mathrm{NV}$ &$|\textrm{RRM}_\mathrm{T}| \lessgtr |\textrm{R}\overline{\textrm{RM}}_\mathrm{T}|$ & $0.0020 $ & $-0.25$  &  $ <0.00001$  \\[5pt]
*$|\Delta \textrm{RM}| $ \& $\pi_\mathrm{NV}$ &  -  &  $|\Delta \textrm{RM}| $ &  $\pi_\mathrm{NV} \lessgtr \bar{\pi}_\mathrm{NV}$ & $<0.00001 $ & $-0.44$  &  $<0.00001 $  \\[5pt]
$I_\mathrm{SP} $ \& $|\log(D)|$ &$\alpha<-0.5$ & $|\log(D)|$& $I_\mathrm{SP}  \lessgtr \bar{I}_\mathrm{SP}  $ &$ 0.40$  & $0.01$  & $ 0.80$  \\[5pt]
*$D$ \& $\alpha$ &  -   & $D$  & $\alpha \lessgtr-0.5$ & $<0.00001$  & $-0.26$  & $< 0.00001$  \\[5pt]
$D$ \& $z$ & $\alpha<-0.5$~\&~$D\ge1.5$ &$D$ & $z \lessgtr \bar{z}$ &$0.015$ &$-0.36$ &$ 0.011$    \\[5pt]
{$I_\mathrm{SP}$ \& $z$} & $\alpha<-0.5$ &$I_\mathrm{SP}$ & $z \lessgtr \bar{z}$ &$ 0.014$ &-0.12 & $0.14 $    \\[5pt]
{$I_\mathrm{SP}$ \& $z$} & $\alpha \ge -0.5$ &$I_\mathrm{SP}$ & $z \lessgtr \bar{z}$ &$0.14 $ &-0.28 & $0.0062 $    \\[5pt]
{$D$ \& $z$} & $\alpha<-0.5$ &$D$ & $z \lessgtr \bar{z}$ &$0.44 $ &-0.03 & $ 0.75$    \\[5pt]
{$D$ \& $z$} & $\alpha \ge -0.5$ &$D$ & $z \lessgtr \bar{z}$ &$ 0.15$ &0.10 & $ 0.36$    \\[5pt]
{$|\textrm{RRM}_\mathrm{T}|$ \& $z$} &  -  &$|\textrm{RRM}_\mathrm{T}|$ & $z \lessgtr \bar{z}$ &$0.074 $ &0.11 & $ 0.10$    \\[5pt]
{$\pi_\mathrm{SP}$ \& $z$} & $\alpha < -0.5$ &$\pi_\mathrm{SP}$ & $z \lessgtr \bar{z}$ &$ 0.73$ &-0.05 & $ 0.51$    \\[5pt]
{$\pi_\mathrm{SP}$ \& $z$} & $\alpha<-0.5 ~\&~ D \ge 1.5$ &$\pi_\mathrm{SP}$ & $z \lessgtr \bar{z}$ &$ 0.79$ &0.12 & $ 0.41$    \\[5pt]
{$\pi_\mathrm{SP}$ \& $z$} & $\alpha \ge -0.5$ &$\pi_\mathrm{SP}$ & $z \lessgtr \bar{z}$ &$0.59 $ &0.01 & $ 0.96$    \\[5pt]
{$\pi_\mathrm{NV}$ \& $z$} & $\alpha<-0.5 ~\&~ D \ge 1.5$ &$\pi_\mathrm{NV}$ & $z \lessgtr \bar{z}$ &$ 0.15$ &0.26 & $ 0.075$    \\[5pt]
{$|\Delta \textrm{RM}|$ \& $z$} &  -  &$|\Delta \textrm{RM}|$ & $z \lessgtr \bar{z}$ &$ 0.47$ &0.04 & $0.55 $    \\[5pt]
*$W1-W2$ \& $\alpha$ &  -   & $W1-W2$ & $\alpha \lessgtr -0.5$ & $<0.00001$ & $0.27$ & $<0.00001$    \\[5pt]
{$W1-W2$ \& $D$} &  -   &$W1-W2$ & $D \lessgtr 0.6$&$ 0.022$ &$ -0.12$&$0.045$     \\[5pt]
{$W1-W2$ \& $D$} &  -   &$W1-W2$ & $0.6<D <1.7$ \& $D>1.7$& $0.31$ & -0.06& $0.38$     \\[5pt]
{$W2-W3$ \& $D$} &  -   &$W2-W3$ & $0.6<D <1.7$ \& $D>1.7$& $0.025$ & 0.06& $0.38$    \\[5pt]
{$W1-W2$ \& $D$} & $\alpha <-0.5$ &$D$ &  $W1-W2 \lessgtr 0.6$ & $0.27$ & $-0.06$ & $0.42$  \\[5pt]
{$W1-W2$ \& $\pi_\mathrm{SP}$} & $\alpha <-0.5$ &$\pi_\mathrm{SP}$ &  $W1-W2 \lessgtr 0.6$ & $ 0.62$ & $-0.07$ & $0.36$  \\[5pt]
\enddata
\tablenotetext{}{Note: The * symbol in the beginning of some of the rows indicates that at least one of the tests resulted in p-value $ \le 10^{-4}$ and p$_\textrm{robust} \le 2\times10^{-3}$. The ? symbol represents suggestive correlations with p-value $ \le 10^{-4}$ and $2\times10^{-3}<$p$_\textrm{robust}< 0.05$. 
}

\end{deluxetable*}

\subsection{WISE colors and polarization}\label{spwise}
We matched our catalog to the Wide-field Infrared Survey Explorer, WISE, catalog, \cite{2010AJ....140.1868W}, with a search radius of five arc-seconds. Out of 533 objects, 455 have WISE counterparts. All of them are detected with at least $5\sigma$ in the WISE $3.4 \mu m$ band, W1, while 445 (323) have $> 5\sigma$ detection in $4.6 \mu m$, W2, ($12 \mu m$, W3) band.  

\begin{figure}[h]
\centering
\includegraphics[scale=0.51]{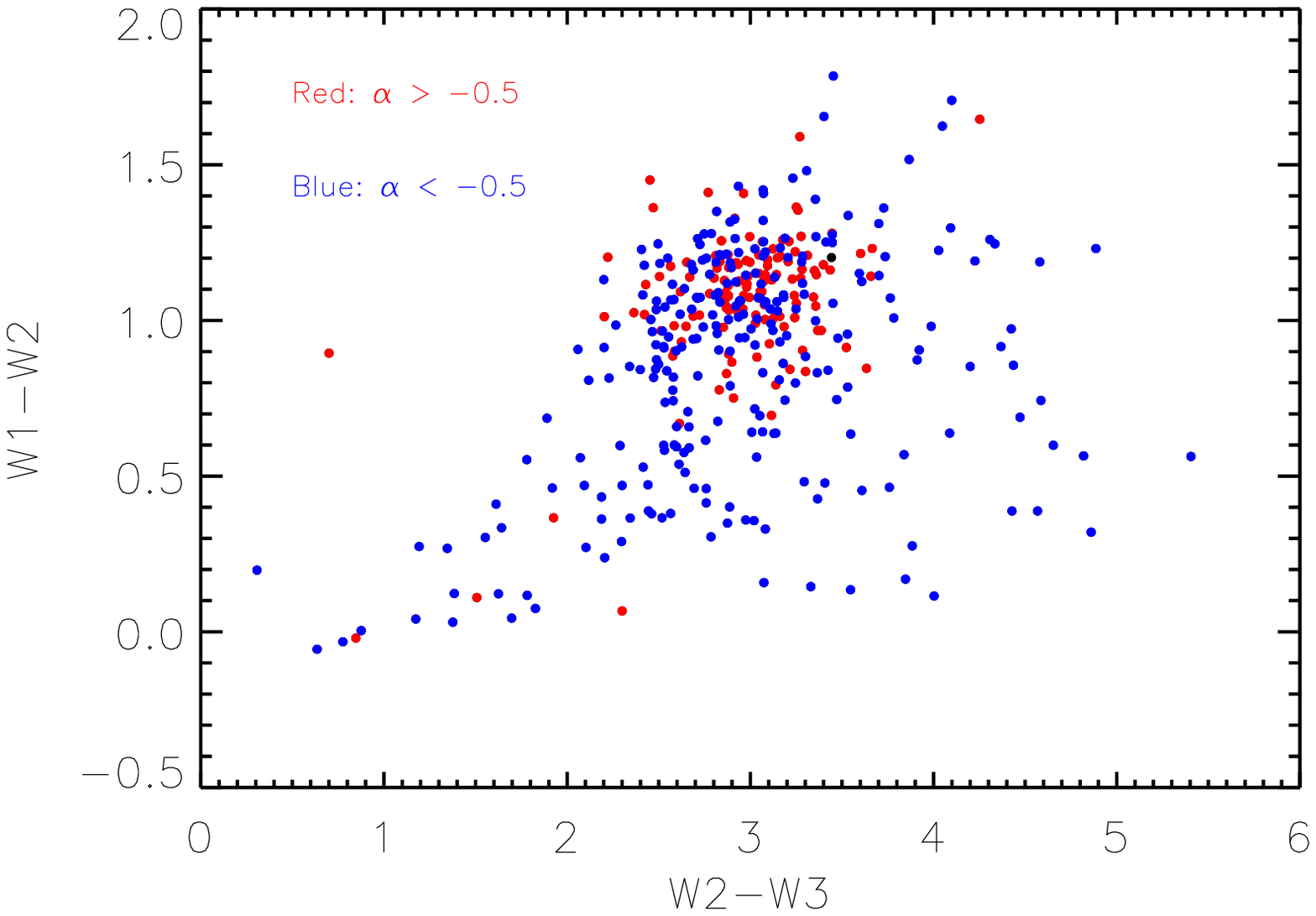}
\caption{Distribution of objects with steep, $\alpha<-0.5$ and flat, $\alpha \ge -0.5$ spectral indices in the WISE color-color diagram. 
\label{fig:wise5arc2}}
\end{figure}

$W1-W2$ and $W2-W3$ colors can be used to separate different galaxy populations such as AGNs and ellipticals \citep{2010AJ....140.1868W,2011ApJ...735..112J}. Recently, \cite{2015MNRAS.453.2326B} studied WISE colors of a large sample of resolved radio galaxies from the Radio Galaxy Zoo project, and found that most radio objects can be classified as ellipticals, AGNs and LIRGs.   
Figure \ref{fig:wise5arc2} shows the WISE color-color diagram of objects in our sample for which we have depolarization measurements and WISE counterparts. All objects used in Figure \ref{fig:wise5arc2} have W1 and W2 detections larger than $5\sigma$ and with small errors in the $W_2-W_3$ colors $\sigma_{(W2-W3)} < 0.4$.  

We investigated the possible dependence of the polarization and depolarization on WISE colors. The WISE dependence is difficult to isolate, since flat and steep spectrum objects have different WISE and different depolarization  distributions.  We therefore looked at steep spectrum objects only, and found that neither $\pi_\mathrm{SP}$ or $|\log(D)|$ were significantly correlated with WISE colors (Table \ref{table2}). We do not sample the ``elliptical'' region of WISE color space, which makes up a distinct population in the  \cite{2015MNRAS.453.2326B} study. 

\subsection{Redshift Dependence}\label{zev} 
\begin{figure}[h]
\centering
\includegraphics[scale=0.51]{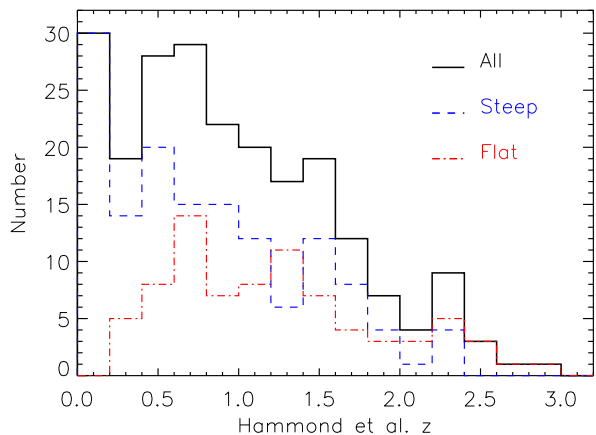}
\caption{Redshift distribution of the matched radio sources with \cite{2012arXiv1209.1438H} catalog. Histograms of steep ($\alpha < -0.5$) and flat ($\alpha \ge -0.5$) spectrum sources are shown in dashed blue and dotted-dashed red lines.  
\label{zhist}}
\end{figure}

There are 222 objects in our sample that are detected in both NVSS and S-PASS polarization maps and have redshifts in  \cite{2012arXiv1209.1438H} catalog. Figure \ref{zhist} shows the redshift distribution of the 222 matched sources, as well as the separated distributions of steep and flat spectrum objects. Steep spectrum objects are located within $0 \le z \le 2.34$ with median redshift of $\bar z=0.64$ while flat spectrum sources, as expected for a flux limited sample, tend to have larger redshifts, $0.22 \le z \le 2.81$, with median of $\bar z=1.18$. 

\begin{figure}[h]
\centering
\includegraphics[scale=0.51]{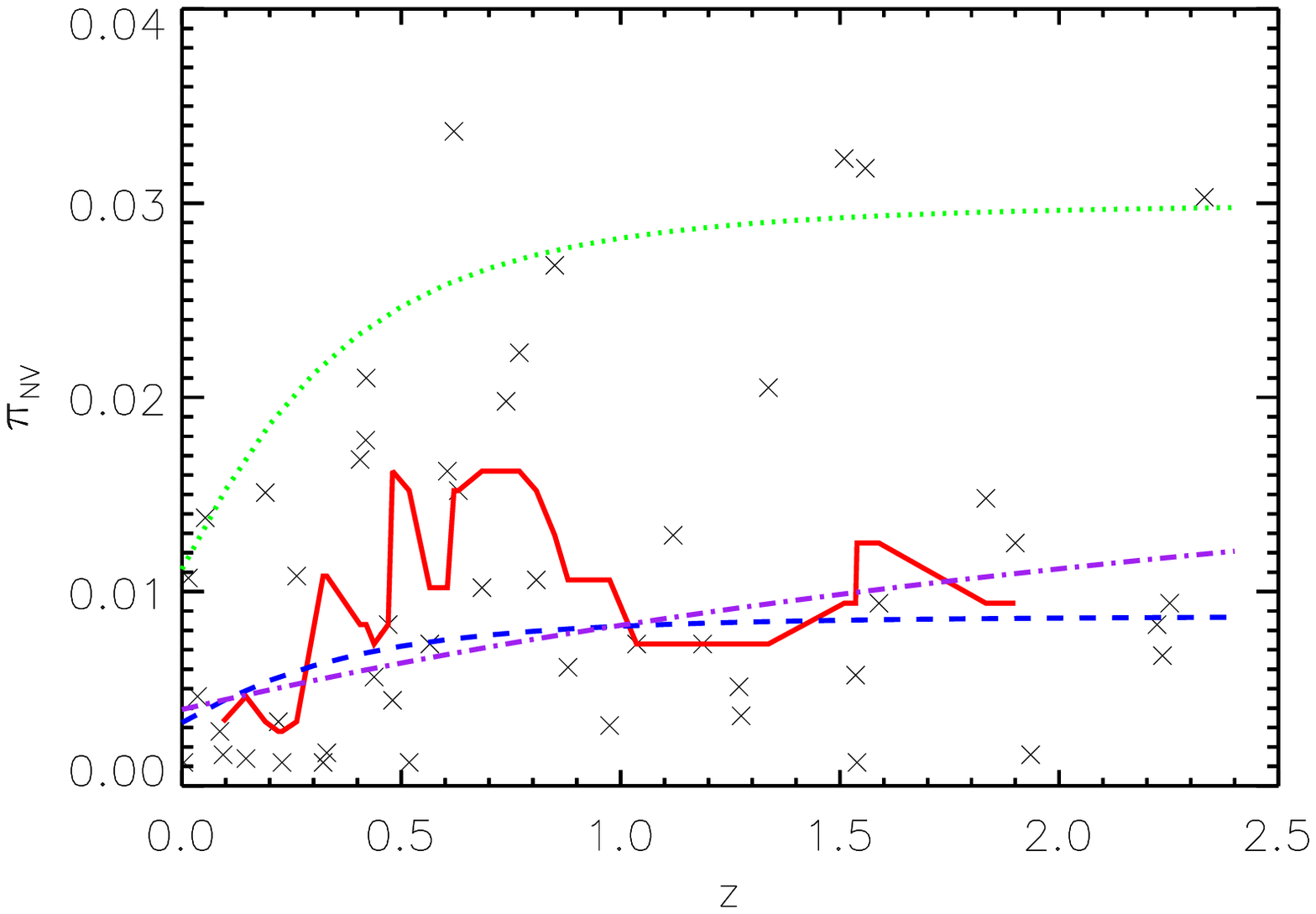}
\includegraphics[scale=0.51]{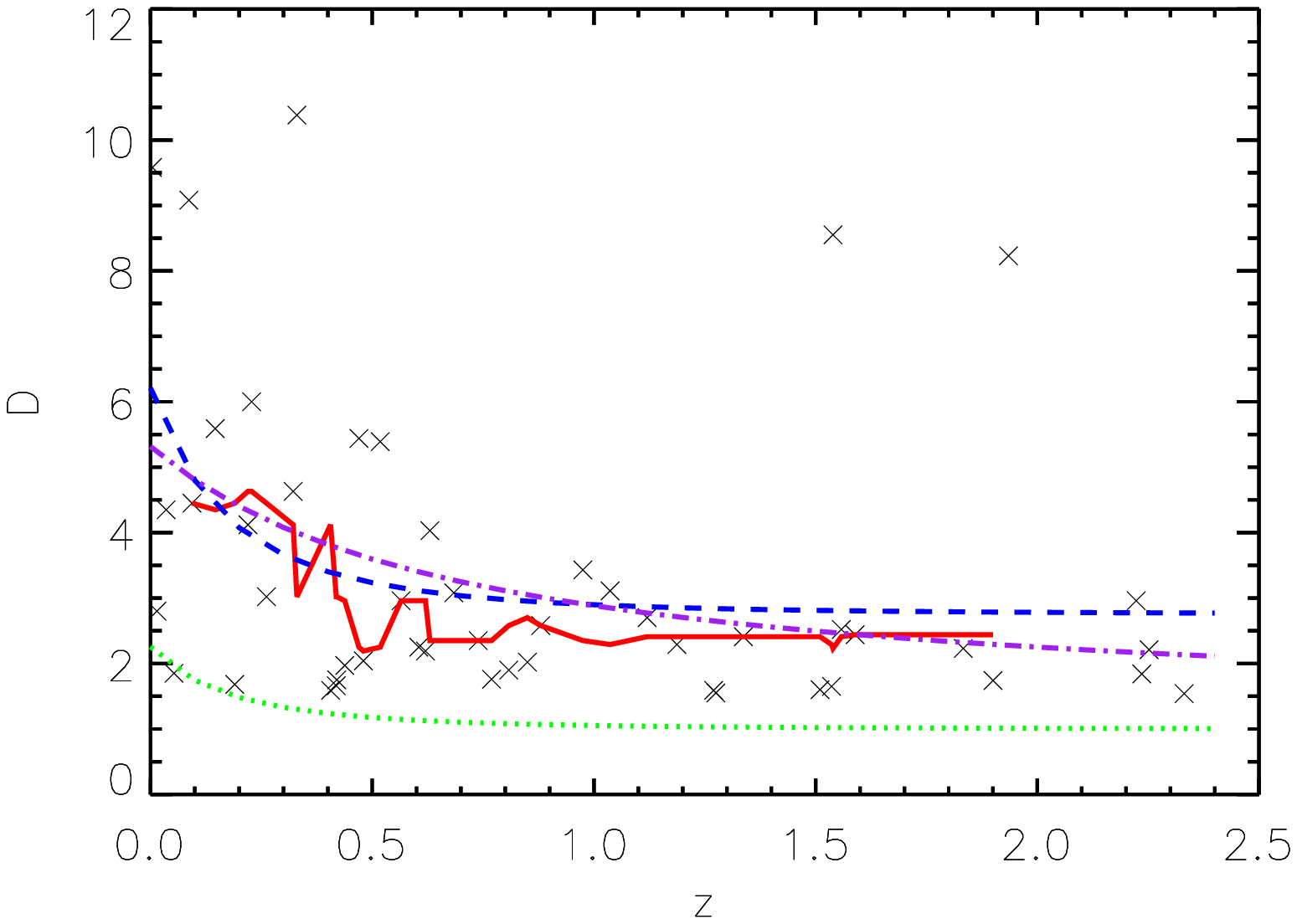}
\caption{Top: Fractional polarizations at 1.4 GHz, $\pi_\mathrm{NV}$, of depolarized steep spectrum sources with $D \ge 1.5$ versus redshift. Bottom: depolarization, $D$, of the same sample of sources versus redshift.  The solid red lines represent the running medians of the $\pi_\mathrm{NV}$ (top) and $D$ (bottom) in bins of redshift. The green dotted, dashed blue and purple dashed-dotted lines are representations of the following three cases with B66 depolarization models: 1. A depolarizing screen located at the redshift of the source,  2. Combination of two depolarizing components, one Galactic and one at the redshift of the source, and 3. An evolving $\sigma_{\phi}$ at the depolarizing screen at the source redshift.    
\label{depolsample}}
\end{figure}

As discussed earlier, steep and flat spectrum objects have different depolarization distributions and therefore, we studied their redshift evolution separately. We examined the redshift dependence of only depolarized steep spectrum sources ($D \ge 1.5$), since we expected to see a change in polarization properties due to the change in rest frame wavelength. We used the threshold $D=1.5$ to choose as many highly depolarized sources as possible while excluding the scattered objects that are in the vicinity of the observed peak at $D\sim1$ in Figure \ref{fig:depfp}. We found weak evidence for a decrease in depolarization of these sources as redshift increases (Spearman $r_s=-0.36$, p=0.011), which does not cross our conservative detection threshold. 
 The average $\pi_\mathrm{NV}$ of 49 steep spectrum sources with $D \ge 1.5$ seems to increase from $\bar\pi_\mathrm{NV}=0.46\%$ at $z \le 0.5$ to $\bar\pi_\mathrm{NV}=1.02\%$ at $z \ge 0.5$, while their observed depolarization decreases and  $\pi_\mathrm{SP}$ stays almost fixed. Figure \ref{depolsample} shows the running median of $\pi_\mathrm{NV}$ and $D$ calculated in bins of redshift as well as the expected evolutionary behavior of the three depolarizing scenarios. We will discuss this more in Section \ref{redshift}.  

\begin{figure}[h]
\centering
\includegraphics[scale=0.51]{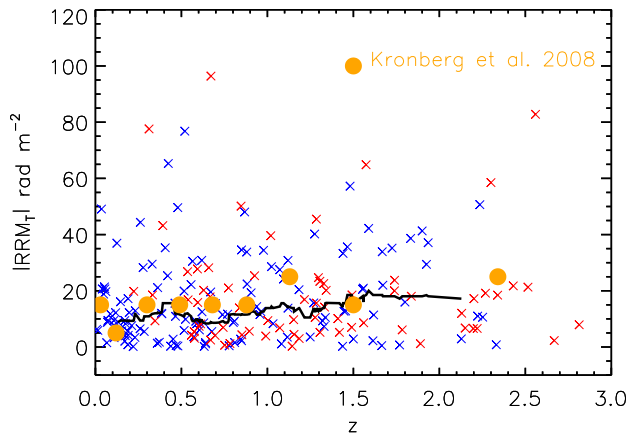}
\caption{ Distribution of the $|\textrm{RRM}|$ for the 206 objects is plotted versus redshift, $z$. Blue and red crosses represent objects with $\alpha <-0.5$ and $\alpha \ge -0.5$. The solid black line shows the running medians of the $|\textrm{RRM}|$ of all sources. The orange filled circles are the data points extracted from Figure 3 of \cite{2008ApJ...676...70K} as discussed in Section \ref{redshift}. Each circle represents the median value of their $|\textrm{RRM}|$ for each redshift bin.  
\label{zrrmt}}
\end{figure}

On the other hand, we do not find any change with redshift in depolarization of separate samples of steep or flat spectrum sources which include all re-polarized and depolarized sources. The median, $\log(D) \approx 0.1$, and standard deviation $\sigma_{\log(D)} \approx 0.26$, of steep spectrum objects stay almost constant with increasing redshift.  Flat spectrum sources appear to be mostly re-polarized at $z<1$ while at higher redshifts the number of re-polarized and depolarized flat spectrum objects are almost the same. However, as listed in Table \ref{table2}, none of the KS and Spearman tests could confirm such a redshift dependence among flat spectrum sources.    
We also performed both KS and Spearman rank tests on $|\textrm{RRM}|$ and $|\Delta \textrm{RM}|$, and did not detect any noticeable redshift dependence (Figure \ref{zrrmt}). The 2.3 GHz fractional polarization of steep and flat spectrum sources also stays fixed at all cosmic times, although have different average values for populations of steep and flat objects. 

\subsection{Summary of major results}
\begin{enumerate}
\item The majority of extragalactic radio objects with $I_\mathrm{SP} \ge 420$ mJy have degrees of polarization on the order of 2\% to 3\% at both 1.4 GHz and 2.3 GHz.
\item $\pi_\mathrm{SP}$ and $|\log(D)|$ are anti-correlated. On average, objects that are not depolarized  ($|\log(D)| \le 0.23$), have median fractional polarizations of $\bar \pi_\mathrm{SP} \approx \bar \pi_\mathrm{NV} \approx 3\% - 4\%$, with $\bar \pi_\mathrm{SP} \approx 2\%$ for more depolarized objects and $\bar \pi_\mathrm{SP} \approx 1\%$ for re-polarized sources. Objects with high fractional polarizations ($\pi_\mathrm{SP}\approx \pi_\mathrm{NV} \approx 10\%$) are not depolarized ($|\log(D)|\approx 0$).
\item Flat and steep spectrum objects have different polarization properties. 55\% of flat spectrum sources are re-polarized, compared to  only 24\% for steep spectrum sources.  Steep spectrum sources have larger degrees of polarization as well as stronger average depolarization.
\item Extended objects ($>20''$) have higher fractional polarizations ($\bar{\pi}_\mathrm{SP}=4\%$) and smaller depolarizations ($|\log(D)| \sim 0.13$) than compact sources ($\bar{\pi}_\mathrm{SP} \sim 2 \%$,  $|\log(D)| \sim 0.20$). 
\item Almost 24\% of the objects with detected polarization have $D > 2$. An additional 10\% of all sources may be too depolarized to be included in our sample. 
\item On average, sources with large $|\log(D)|$ (depolarized or re-polarized) show larger  
changes in $\textrm{RM}$ with wavelength ($\Delta \textrm{RM}$). 
\item We find weak evidence for a redshift dependence of the depolarization in a sub-sample of sources, those with steep spectra and  $D\ge1.5 $.   
\item We do not find any evidence for changes of the observed 2.3 GHz  fractional polarization, depolarization, $|\textrm{RRM}_\mathrm{T}|$ and $\Delta \textrm{RM}$  from $z=0$ to $z=2$ when all sources are considered. The median degree of polarization of both steep (141) and flat (81) spectrum sources with known redshift remain almost constant at $\pi_\mathrm{SP} \approx 2.5 \%$ and  $\pi_\mathrm{SP} \approx 2.0 \%$ respectively.
\item A large scatter in both depolarization and fractional polarization is seen at all redshifts.  
\item We did not find any evidence for angular clustering in the distribution of the depolarized sources.
\item Both  $\pi$ and $|\log(D)|$ of steep spectrum sources are independent of WISE $W_1-W_2$ color.

\end{enumerate}

\section{Discussion}\label{discussion}

\subsection{Radio source field disorder}\label{obs} 
While radio synchrotron radiation can potentially be highly polarized, the NVSS and S-PASS fractional polarizations of most objects in our sample are around $2\% - 3\%$, and very rarely exceed $10\%$ (Figure \ref{fig:fphist}).   
Depolarization due to the presence of an irregular Faraday screen between the source and the observer, e.g.,  can potentially reduce the initial degree of the polarization, generally leading to higher fractional polarizations at higher frequencies \citep{1966MNRAS.133...67B,1991MNRAS.250..726T}. However, between 1.4 GHz and 2.3 GHz we find that the majority of extragalactic objects experience only small depolarizations, with $60\%$ of the objects have $0.6< D < 1.7$.  Moreover, objects with the strongest fractional polarizations ($\pi\approx 10\%$) have little depolarization.  The reduction from a theoretical maximum of $\sim$40-70\% to  either $\approx$10\% with no depolarization, or $\approx$3\%, with modest depolarization, must therefore be due to field disorder.

To approximate the necessary number of randomly oriented magnetic field patches within an unresolved source, we performed a simple simulation. We considered a uniform brightness two dimensional source, with equal fractional polarizations $\pi_0=50\%$ in each patch. By randomizing the polarization angles, we estimated that sources currently unresolved in our beam should contain approximately 70 to 80 independent magnetic patches to reduce the observed fractional polarization to $\sim$4\%.

There is a subset of sources where depolarization does play a significant role. Almost, 24\% of sources with detected polarizations have $D>2$. Moreover, we estimated a missing $\approx$10\% population of heavily depolarized sources. It is not clear how strong an effect field disorder has for that subset.

\subsection{Prospects for high frequency surveys}
One important implication of these results is for surveys at higher frequencies, where one might expect to increase number counts by a large factor because of less depolarization.  However, changing the frequency of observation from L to S band will not result in a major increase in the number of polarized detections. the number of polarized objects. As an example, the number of sources with polarized flux densities larger than 10 mJy in our sample is almost equal at both 2.3 GHz and 1.4 GHz (368 in S band and 363 in L band). Future polarization surveys and the Square Kilometer Array, SKA \citep{2011arXiv1111.5802B} precursors such as Polarization Sky Survey of the Universe's Magnetism,  POSSUM \citep{2010AAS...21547013G}, Westerbork Observations of the Deep APERTIF Northern sky, WODAN \citep{2012MNRAS.427.2079C}, MeerKAT International GigaHertz Tiered Extragalactic Exploration survey, MIGHTEE \citep{2012AfrSk..16...44J}, Very Large Array Sky Survey, VLASS \citep{2014arXiv1401.1875M} and VLASS Deep will detect hundreds of thousands of polarized sources in different frequencies. The VLASS will operate at S band from 2 to 4 GHz and has angular resolution and sensitivity of $\sim 3.5$ arcsec and 0.7 mJy per beam respectively. The number density of flat spectrum sources is expected to be similar in L and S bands since their flux density is almost independent of the frequency, and their median depolarization is $\bar D \sim 1$ as shown in Figure \ref{fig:depol}. On the other hand, steep spectrum, $\alpha < -0.5$, sources in our sample with median $\bar{\alpha}=-0.9$ are on average fainter at S band by a factor of 1.4. Therefore, their number density at a fixed signal to noise reduces. However, the median polarization of steep spectrum objects in our sample is approximately 1.3 times higher at 2.3 GHz than 1.4 GHz at resolutions as low as S-PASS, $\sim 9$ arcmin. This indicates that the median polarization flux density of these objects should have been reduced by $\sim$17\%.  \cite{2014ApJ...785...45R} showed at 1.6 arcsec resolution there are $\sim 6$ polarized sources per squared degree at 0.7 mJy per beam and S:N $>$ 10 in L band, and the integrated number density of objects with polarization flux density larger than $p$ goes as $N_p\propto p^{-0.6} $.  As a result, one can expect to detect roughly 11\% less polarized objects at S band compared to L band at 1.6 arcsec resolution. All in all, considering the larger beam size of the VLASS all sky survey one can expect to detect approximately the same number of polarized sources in S band as the calculation of \cite{2014ApJ...785...45R} in L band. This is already a factor of six above the existing surface density of polarized sources from the NVSS catalog in L band. 

\subsection{Prospects for $\textrm{RM}$ grid experiments}
There is strong interest in measuring and estimating the intergalactic magnetic field in clusters of galaxies or in cosmic filaments through $\textrm{RM}$ analysis and tomography, e.g.  \cite{2014PASJ...66...65A}. In the presence of a single Faraday screen along the line of sight, the rotation angle of the radio polarization vector of extra-galactic sources depends linearly on $\lambda^2$. This simple relation makes it possible to estimate the magnetic field of the medium with some assumptions for the electron density, after subtracting out a Galactic component. However, any complication in the structure of the Faraday screen within the observation beam or along the line of sight through the emitting source will result in non-$\lambda^2$ behavior, and an inability to isolate the foreground screen of interest. 

 We have measured the non-$\lambda^2$ behavior using $\Delta \textrm{RM}$.  As shown in Figure \ref{drmp}, large $\Delta \textrm{RM}$s occur preferentially at low fractional polarizations. In order to avoid large values of $\Delta \textrm{RM}$, which would compromise any foreground experiment, it is necessary to use only fractional polarizations ($\ge 3-4\%$).  This will cause a reduction in the number of available sources;  only 33\% of sources in our sample have $\pi_{SP} >$ 3\%. However, if reliable $\chi(\lambda^2)$ were available for some subset of sources, then it might be possible to increase this number.

\subsection{Origins of depolarization}

As shown in Section \ref{galD} we did not detect any angular clustering of sources by fractional polarization or depolarization, that would have implied a Galactic origin.  We can not rule out the possibility of Galactic $\textrm{RM}$ fluctuations on arcsec scales, but these are likely to be extremely small and we do not consider them further here. 

The dependence of depolarization on spectral index shows that it must primarily occur local to the source. If depolarization is local to the environment of the source, then it may show signs of dependence to some intrinsic characteristics of the source such as spectral index or the luminosity. The results found here on the spectral behavior are consistent with \cite{2014ApJS..212...15F} who did a multi-wavelength polarization study on sources selected from the TSS09 catalog.

The dependence of polarization properties of objects on their angular extent (Section \ref{extent}) also supports the local depolarization scenario. As shown in Figure \ref{fig:dsize}, compact sources seem to have larger depolarizations ( $|\log(D)| \sim 0.20$ vs. $\sim 0.13$) and smaller fractional polarizations ($\bar{\pi}_\mathrm{SP}=4\%$ vs. 2\%) than sources extended in NVSS.  This is inconsistent with irregular screens either Galactic or extragalactic, which should yield higher fractional polarizations and less depolarization for compact sources.  Thus, the depolarization must arise in a Faraday component directly related to the source. If Galactic or intervening Faraday screens were the dominant depolarizing components then we expect to see larger depolarization in a sample of extended sources. 
 
\subsubsection{The origin of the total intensity and fractional polarization anti-correlation}\label{origin}
The anti-correlation between total intensity and fractional polarization at 1.4 GHz has been extensively discussed (such as \citealt{2002A&A...396..463M, 2004MNRAS.349.1267T, 2007ApJ...666..201T, 2010ApJ...714.1689G,2010MNRAS.402.2792S,2014ApJ...787...99S}). Recently, \cite{2014MNRAS.444..700B}  used WISE colors to suggest that the anti-correlation was due to the difference in environments between WISE-AGNs (IR colors dominated by AGN) and WISE-Ellipticals (IR colors dominated by starlight).     
These effects are likely confused by the fact that the anti-correlation is found only among steep-spectrum sources, as discussed in Section \ref{ip}. The WISE-AGN class contains a large fraction of flat spectrum objects, for which we find no anti-correlation, while the WISE-Ellipticals are largely steep-spectrum \citep{2014MNRAS.444..700B}.  The dependence we found on the spectral index is also consistent with \cite{2004MNRAS.349.1267T} and the stacking analysis of \cite{2015arXiv150100390S}.

The limited range of $I_\textrm{SP}$ in our sample makes it difficult to study these effects. However, to illuminate the underlying issues, we note that the suggestive anti-correlation between $I_\mathrm{SP}$ and $\pi_\mathrm{SP}$ of steep spectrum sources must arise from some physical difference in properties between the bright and faint sources that are not expected in fair, uniform samples.    We have not been able to identify this underlying parameter.  We find no statistically significant anti-correlation between $L_\mathrm{SP}$ and  $\pi_\mathrm{SP}$.  We attempted to correct for the size dependence, in case that was a confounding variable, but the anti-correlation remained. Size could still be an important factor, since the resolution of even the NVSS is much larger than the typical source size.  Higher resolution observations of this sample could reveal, e.g., that the bright sources are much more compact and dominated by central AGN, as opposed to fainter, lobe-dominated structures with more ordered fields.   

Depolarization might also be playing a role, since $\pi_\mathrm{SP}$ is correlated with the $|\log(D)|$. However, again, the anti-correlation breaks down when we look at   $L_\mathrm{SP}$ and $|\log(D)|$.  This leaves us back, again, at some as yet undetermined physical difference between the faint and bright sources.  

\subsubsection{Re-polarized objects}
We showed that most re-polarized objects have flat spectra ($\alpha \ge -0.5$), and are therefore concentrated in the WISE-AGN population (Figure \ref{fig:wise5arc2}).  This makes it likely that they contain a high proportion of compact nuclei with polarization SEDs influenced by self-absorbed, and perhaps Faraday thick components.   This is consistent with \cite{2014ApJS..212...15F} who also found flat spectrum objects have complex polarization behaviors.  

While 61\% of re-polarized objects have flat spectra and are optically thick sources, the remaining 39\% have steep spectra. The nature of these objects is not clear. However, there are few proposed models in the literature. re-polarization can occur when there is interference between two (or a few) unresolved and separate Faraday patches in the beam of the telescope. This can result in an oscillatory behavior of the fractional polarization with changing frequency as discussed in \cite{2011AJ....141..191F} and \cite{1984ApJ...283..540G}.\cite{2012AJ....144..105H} studied the AGN jet structure of 191 extragalactic radio objects, and found multiple regions along the jets of a few objects show signs of re-polarization. As discussed in \cite{2012ApJ...747L..24H} they argue that both internal Faraday rotation in the jet medium as well as the configuration of the magnetic fields can explain the observed re-polarization in these optically thin jets. In Faraday thick regions the rotation of the polarization angles might align the polarization vectors from the far and near sides along the line of sight which can potentially result in re-polarization.  

\subsection{Redshift Evolution}\label{redshift}
The evolution of the magnetic properties of galaxies with time has been subject of multiple studies (such as \citealt{2012arXiv1209.1438H, 2008ApJ...676...70K,2008Natur.454..302B,2005MNRAS.359.1456G,1995ApJ...445..624O,1984ApJ...279...19W}). We distinguish here between two different quantities, an \emph{observed} redshift dependence and an \emph{inferred} redshift evolution, based on applying the polarization equivalent of a K-correction (redshift dilution).   

As discussed in Section \ref{zev}, we found weak evidence that the average observed depolarization of steep spectrum depolarized sources with $D \ge 1.5$ decreases with increasing redshift, while the 1.4~GHz fractional polarization increases (the 2.3~GHz fractional polarization shows no change). The detected redshift variations are weak, compared to the scatter, and their probability (0.011) does not cross our conservative detection threshold.  However, given the importance of this issue, we discuss the causes and consequences of redshift dependencies to help clarify the underlying issues.

Polarization SEDs are often complex, especially for flat spectrum sources.  This is seen in our numerous detections of re-polarization, and the broad wavelength SEDs cataloged by \cite{2014ApJS..212...15F}.  In such cases, it is impossible to predict the trends of depolarization and fractional polarization with redshift expected from the K-correction.  In the case where $D \sim 1$, no redshift dependence is expected, since there is no wavelength dependence to the fractional polarization.  Therefore, the fact that we observe decreasing depolarization and increasing 1.4~GHz fractional polarization at increased redshift only for steep-spectrum sources with $D > 1.5$ is consistent with K-corrections only, without any physical redshift evolution.

We now look at this more quantitatively, assuming the simplest case of an unresolved source with an irregular depolarizing Faraday screen  \citep{1966MNRAS.133...67B}(B66), external to, but at the same redshift as the source. 
The expected fractional polarization behavior is then
\begin{equation}
\label{B66}
\pi=\pi_0 \exp(-C\lambda_{rest}^4)
\end{equation}
where $\pi_0$ is the initial fractional polarization and $C \propto \sigma_{\phi}^2$ is a function of the dispersion in the Faraday depth. For a region with electron density $n$ and magnetic field component parallel to the line of sight $B_z$, fluctuations in the parameter $nB_z$ over the extent of the region is represented by $\sigma_{\phi}$. Assuming no physical change in $\sigma_{\phi}$ with time, the redshift dilution effect results in an increase in the observed fractional polarization, $\pi\propto exp\left(-C\lambda^4(1+z)^{-4}\right)$.  The observed depolarization also decreases with redshift since $D\propto exp\left(C(\lambda_\mathrm{NV}^4-\lambda_\mathrm{SP}^4)(1+z)^{-4}\right)$.  This simplest picture (Model 1), however, is not quantitatively consistent with our observations (Figure \ref{depolsample}).

We therefore considered two additional models based on the B66 screen. Model 2:  A combination of two depolarizing components, one Galactic or relatively local to us,  and one at the redshift of the source, and Model 3: A physical change in  $\sigma_{\phi}$ of the depolarizing screen at the source redshift. 
As shown in Figure \ref{depolsample}, the general behavior of the observed  $\pi_\mathrm{NV}$, and $D$ as well as $\pi_\mathrm{SP}$ (not shown) of the depolarized steep spectrum sources and their evolution with redshift can be explained by models 2 and 3. However, a single depolarizing component, local to the source, with no evolution in $\sigma_{\phi}$ does not seem to be consistent with the observation. Larger samples, and resolved polarization maps where the Faraday structure can be directly seen,  are needed to clarify these results. 

As an alternative to the B66 screen,  \cite{1991MNRAS.250..726T} suggested depolarization can be modeled as power law $\pi \propto \lambda^{-4/m}$ at wavelengths larger than $\lambda_{1/2}$, at which the degree of polarization is equal to  half of its maximum value. The above relation only holds under certain condition in which the Faraday screen $\textrm{RM}$ structure function varies as a power law across the source $S(\delta x)  \propto \delta x^m$ where $S(\delta x)\equiv <[\textrm{RM}(x+\delta x)-\textrm{RM}(x)]^2>$ and $x$ is the angular coordinate. If we assume the fractional polarization of unresolved objects follows any power law model with arbitrary exponent $-4/m$ and a constant related to the $\textrm{RM}$ dispersion, $\pi =C \lambda^{-4/m}$, then the observed depolarization, $D=\pi_{SP}/\pi_{NV}$, and both the redshift and the $\sigma_{\phi}$ dependences cancel out. Therefore, one can expect to observe no evolution in the average $D$ even if $\sigma_{\phi}$ changes with redshift, contrary to what we observe.
  
\subsubsection{Comparisons to previous work}
Earlier work has been based on samples including sources with both flat and steep spectra, and without selections based on depolarization.   For our full sample, we find no redshift trends in fractional polarizations or depolarization, or measures of increased Faraday structure such as $|\textrm{RRM}_\mathrm{T}|$ and $|\Delta \textrm{RM}|$.  This is consistent with the negative results from \cite{2012ApJ...761..144B} and \cite{2012arXiv1209.1438H}.  In addition, their samples were taken from the TSS09 catalog, which is biased towards high fractional polarizations, and thus, towards depolarizations D$\sim$1, for which no redshift evolution is expected. 

Our data are inconsistent with the analysis of \cite{2008ApJ...676...70K},  who claimed that the rotation measure of galaxies at redshifts larger than $z=1$ are on average larger (by $\sim 10$ rad m$^{-2}$) than the low redshift objects, despite the redshift dilution effect. 
In Figure \ref{zrrmt} we show $|\textrm{RRM}_\mathrm{T}|$ versus the redshift of objects in our sample and overlay the  \cite{2008ApJ...676...70K} median $|\textrm{RRM}|$ values from their Figure 3.  Our data are consistent with theirs, and show no evidence for the claimed increase in $\textrm{RRM}$.

It is possible that a physical increase in $\sigma_{\phi}$ and depolarization as a function of redshift could mask the redshift dilution effect, leaving no observed redshift dependence to fractional polarization, $\textrm{RRM}$, $\Delta \textrm{RM}$ or depolarization.  
 This is discussed with more details in \cite{2012arXiv1209.1438H}, \cite{2008ApJ...676...70K}, \cite{2008Natur.454..302B}, \cite{1995ApJ...445..624O} and \cite{1984ApJ...279...19W}. 
 
\cite{2005MNRAS.359.1456G} studied the redshift evolution of the depolarization of 26 resolved,  powerful radio galaxies and quasars over the cosmic time. They applied corrections to the measured depolarizations based on models of the wavelength and resolution effects at different redshifts. They claim a physical evolution in $\sigma_{\phi}$ and depolarization as a function of redshift, but we cannot compare their results to ours, since neither the original data nor the details of the models are shown. 

\section{Conclusions} \label{summary}
We constructed a depolarization ($D=\pi_{2.3}/\pi_{1.4}$) catalog of extragalactic radio sources brighter than $420$ mJy at 2.3 GHz including total intensities, spectral indices, observed and residual rotation measures, fractional polarization, depolarization as well as the redshift, 2.3 GHz luminosity and WISE magnitudes for almost half of the objects. We looked for possible correlations between these quantities and found that the fractional polarization of extragalactic radio sources depends on the spectral index, morphology, the intrinsic magnetic field disorder as well as the depolarization of these sources. We summarize our main conclusions as follows: \\

Consistent with previous studies over half of flat spectrum sources in our sample are re-polarized while the majority of steep spectrum objects are depolarized. There is also a significant population of steep-spectrum sources that are repolarized; their underlying physical structure is currently unknown.  Although steep objects are more polarized at 2.3~GHz, they are fainter in total intensity, and therefore future surveys at higher frequencies will result in approximately the same number of sources at fixed sensitivity as the lower frequencies. 

Depolarization, and thus fractional polarizations, are related to the presence of Faraday structures indicated by
the non-$\lambda^2$ behavior of polarization angles ($\Delta \textrm{RM}$). 
Future studies using polarized sources as background probes need to minimize $\textrm{RM}$ structures intrinsic to the sources.  Such clean samples require high fractional polarizations ($\pi \ge 4\%$), which will severely limit the number of available sources.  

Sources with little or no depolarization between 1.4 GHz and 2.3 GHz  have fractional polarizations ranging from a few to 10\%. This is much lower than the theoretical maximum, and therefore shows the dominant role of field disorder in creating low polarizations.  
Compact steep spectrum objects in the NVSS catalog have more Faraday structure, and are $\sim 2$ times less polarized at 2.3 GHz than the extended sources.

 We found suggestive evidence for a decrease in the depolarization from $z=0$ to $z=2.3$, but only when the sample is restricted to the steep spectrum, $\alpha < -0.5$, depolarized, $D \ge 1.5$ objects. More investigation is needed to confirm the depolarization trend. Assuming that it's real, it is likely the result of the redshift dilution effect (at least partially) but requires more than a simple depolarizing screen local to the source.\\

The National Radio Astronomy Observatory is a facility of the National Science Foundation operated under cooperative agreement by Associated Universities, Inc. Partial support for ML and LR comes from National Science Foundation grant AST-1211595 to the University of Minnesota. B.M.G. has been supported by the Australian Research Council through the Centre for All-sky Astrophysics (grant CE110001020) and through an Australian Laureate Fellowship (grant FL100100114). The Dunlap Institute is funded through an endowment established by the David Dunlap family and the University of Toronto. We would like to thank G. Bernardi and D. H. F. M. Schnitzeler and the referee for a number of useful conversations and comments on the manuscript.

\bibliographystyle{apj}
\bibliography{ref}

\end{document}